\documentclass{article} % For LaTeX2e

\usepackage{graphicx} 
\usepackage[utf8]{inputenc} % allow utf-8 input
\usepackage[T1]{fontenc}    % use 8-bit T1 fonts
\usepackage{url}            % simple URL typesetting
\usepackage{booktabs}       % professional-quality tables
\usepackage{amsfonts}       % blackboard math symbols
\usepackage{nicefrac}       % compact symbols for 1/2, etc.
\usepackage{microtype}      % microtypography
\usepackage{xcolor}         % colors
\usepackage{amsmath,amsfonts,bm}

\def\eqref#1{equation~\ref{#1}}
\def\1{\bm{1}}

\def\vtheta{{\bm{\theta}}}
\def\va{{\bm{a}}}
\def\vb{{\bm{b}}}

\def\ve{{\bm{e}}}

\def\vg{{\bm{g}}}
\def\vh{{\bm{h}}}

\def\vm{{\bm{m}}}

\def\vp{{\bm{p}}}

\def\vs{{\bm{s}}}

\def\vu{{\bm{u}}}

\def\vw{{\bm{w}}}
\def\vx{{\bm{x}}}
\def\vy{{\bm{y}}}
\def\vz{{\bm{z}}}

\def\v1{{\bm{1}}}

\def\mA{{\bm{A}}}

\def\mC{{\bm{C}}}

\def\mE{{\bm{E}}}

\def\mP{{\bm{P}}}
\def\mQ{{\bm{Q}}}

\def\mS{{\bm{S}}}

\def\mW{{\bm{W}}}
\def\mX{{\bm{X}}}

\def\mZ{{\bm{Z}}}

\DeclareMathAlphabet{\mathsfit}{\encodingdefault}{\sfdefault}{m}{sl}
\SetMathAlphabet{\mathsfit}{bold}{\encodingdefault}{\sfdefault}{bx}{n}

\DeclareMathOperator{\Tr}{Tr}

\usepackage{hyperref}       % hyperlinks
\usepackage{zref-clever}
\zcsetup{abbrev}
\usepackage{dsfont}
\usepackage{comment}
\newcommand{\will}[1]{\textcolor{blue}{Will: #1}}
\newcommand{\cref}[1]{\zcref{#1}}
\newcommand{\Cref}[1]{\zcref[S]{#1}}
\zcRefTypeSetup{appendix}{Name-sg-ab=App.,name-sg-ab=app.,Name-pl-ab=Apps.,name-pl-ab=apps.}

\usepackage{iclr2027_conference,times}

\title{\centering{how optimality structures sparse dictionaries \\
theory for interpreting sae representations}}

\author{%
  William Dorrell \\
  Kempner Institute,
  Harvard University
}

\iclrfinalcopy % Uncomment for camera-ready version, but NOT for submission.
\begin{document}

\maketitle

\begin{abstract}
Sparse Autoencoders (SAEs) have found success parsing neural network representations into interpretable concepts, providing a basis for understanding and control. However, what exactly SAEs extract and, hence, the scientific conclusions we can draw from them are not obvious. In short, if your SAE behaves strangely, does that reflect interesting neural network behaviour or an SAE-imposed distortion? Towards answering this, we use dictionary learning identifiability results to derive constraints that optimal dictionary learning features must satisfy. For example, an optimal feature will never turn on only while another is active. We use these conditions to explain various SAE oddities - hierarchical splitting \& absorption, which features can be left in the residuals, dense antipodal features, and infinite feature splitting - simply as properties imposed by the dictionary learning objective. Finally, these constraints are diagnostic: real SAEs pass when measured on the dataset on which they were trained, but increasingly fail as the test dataset becomes more `distant'. In sum, we hope to provide theoretical tools to explain puzzling SAE patterns, allowing more principled inferences about internal model behaviour.

\end{abstract}

\section{introduction}

Dictionary learning methods model data using a sparse linear combination of learnt components. They offer practical algorithms that operationalise the intuition that the world is composed of parts (a chair, the sun, twilight, ...) each contributing to a small portion of experience, and learns these parts from observations. They have found widespread use, both practically \citep{mairal2008sparse,wright2008robust,elad2010sparse,dumitrescu2018dictionary}, and as a normative model of brains' sensory processing \citep{olshausen1996emergence,olshausen2004sparse,sadagopan2009nonlinear,willmore2009auditory}. 

More recently, mechanistic interpretability, which tries to understand neural networks as a neuroscientist does the brain, has driven a surge of interest in dictionary learning as a tool to understand neural networks. This was motivated by the finding that neural networks often encode interpretable concepts as linear directions \citep{mikolov2013linguistic,arora2016latent}. Further, these concepts are often sparse, and there appear to be more of them than the representation has neurons, forcing the network to encode them in `superposition' \citep{arora2018linear,elhage2022toy}. While superposition can hinder single neuron interpretation, conceptualising representations as sparse linear combinations of concepts - the `linear representation hypothesis' - naturally motivates using dictionary learning to recover the concepts. Indeed, flavours of Sparse Autoencoders (SAEs), a dictionary learning variant, have proven effective at extracting meaningful concepts from LLM representations \citep{bricken2023towards,cunningham2023sparse,templeton2024scaling}, and these concepts have proved capable of controlling network output \citep{templeton2024scaling,durmus2024evaluating,arad2025saes}.

What can the success of SAEs teach us about how neural networks represent and manipulate information? At first blush, it seems a ringing endorsement for the `linear representation hypothesis': one can find linearly encoded concepts that control the network; this surely reflects models' internal use of these same vectors? However, a growing body of empirical work has produced puzzling observations. For example, training SAEs of increasing size (i.e. with more and more concepts) leads to `feature splitting': a concept identified by a smaller SAE, such as `base64', splits into multiple in larger SAE, like `digits in base64', `letters in base64' and `ASCII characters in base64' \citep{bricken2023towards}. This is perhaps surprising; a hierarchical decomposition in which `base64' and `ASCII in base64' coexist seems natural. Is feature splitting evidence that neural networks don't encode hierarchies?

Thankfully, the answer is likely `no'. %For example, newer concept extraction methods built with different inductive biases do indeed respect such hierarchical relationships \citep{bussmann2025learning,costa2025flat}. 
As we shall discuss, dictionary learning cannot help splitting hierarchies. This illustrates our motivation: dictionary learning assumes representations can be well-modelled as sparse linear combinations of parts. When we apply dictionary learning techniques, like SAEs, to neural networks we are not just seeing symptoms of neural network behaviour - but also the assumptions baked into our tools. Further, the way those assumptions manifest is often surprising. As such, a better understanding of our tools will allow us to more accurately link our observations to network behaviour, and provide targets for developing tools that more precisely reflect our goals.

Here, we take steps towards such an understanding. Our main result is conditions that a representation must obey to be a dictionary learning optima, analogous to earlier work in slightly different settings \citep{gribonval2010dictionary}. These conditions are interpretable, describing how optimal features and residuals are allowed to relate to one another. We use these constraints to explain SAE behaviours - hierarchical splitting \& absorption, the structure of residuals, and dense antipodal features. Additionally, we highlight a wide-convex reformulation of the dictionary learning problem which shows that widening SAEs will refine their categorisations until each ray of data is assigned a concept \citep{bach2008convex}. Finally, we show our conditions are diagnostic: real SAEs pass when measured on the dataset on their training set, but progressively fail for more distant datasets. In sum, we provide tools for understanding the behaviour of dictionary learning, and use it to relate otherwise puzzling SAE properties to the interaction between data and modelling assumptions. In so doing, we hope to motivate alternative interpretability approaches that reflect more realistic representational structures.

\section{related work: 3 related threads}\label{sec:lit_review}

\textbf{Dictionary Learning Identifiability} There has been extensive work on the identifiability of dictionary learning, drawing on earlier work in compressed sensing (sparse coding with a fixed dictionary) \citep{candes2006near,donoho2006compressed,donoho2003optimally}, and matrix factorisation \citep{tatli2021polytopic,gillis2020nonnegative}. Much of it assumes a particular data-generating process, then describes conditions under which the true factors are recovered \citep{geng2014local,schnass2015local,hillar2015can,wu2018local,gribonval2015sparse,cohen2019identifiability,wang2020unique}. Here, motivated by the heterogeneity of neural network representations, we focus on approaches that guarantee optimality without data-generating assumptions. There exist global identifiability results of this generality \citep{hu2023global}, but they require novel formulations of the problem. For more standard dictionary learning problems \citet{gribonval2010dictionary} derive local optimality conditions without data assumptions, and we use these techniques in our work. Further, we use the results of \citet{bach2008convex} to reformulate dictionary learning and study the large width limit. Relative to this literature our novelty lies in interpreting a set of conditions similar to those of \citet{gribonval2010dictionary} and making them do work to explain the behaviour of SAEs.

\textbf{SAE Theory} We highlight three theoretical approaches to understanding SAE representations. First, two recent works study the conditions under which SAEs recover sparse features: \citet{cui2026limits} show recovery under extreme sparsity of features whose embedding dot-products are nonpositive, while \citet{tang2025theoretical} study an approximate loss valid under sufficient sparsity and show that it is minimised only by true feature recovery. By making specific dataset assumptions, these are analogues of older dictionary learning work. We instead study constraints on generic dictionary learning optima allowing us to study the representation of data that doesn't respect sparse data-generating assumptions. Second, \citet{hindupur2025projecting} study how architectural constraints (e.g. choosing TopK or JumpReLU SAE \citep{gao2024scaling,rajamanoharan2024jumping})  determine the concepts SAEs can extract. By studying vanilla dictionary learning our work ignores such architectural effects. Instead we focus on the distributional assumptions in SAEs, a relatively under-explored aspect. Third, follow-up work on feature splitting and absorption linked these phenomena to the interplay of sparsity and hierarchy \citep{till_2024_saes_true_features,chanin2024absorption}. We build on these works, making the same link more robustly.

\textbf{Neural Coding} There's decades of neuroscientific work relating representations to solutions of optimisation problems \citep{laughlin1981simple,atick1990towards,simoncelli2001natural}, including, famously, dictionary learning \citep{olshausen1996emergence}. We draw particular technical inspiration from \citet{sengupta2018manifold, dorrell2026convex} who show the convexity of nonnegative neural coding problems by lifting to the space of representational similarity matrices. This is a particular example of the semi-definite relaxation, common in polynomial optimisation \citep{shor1987class,lasserre2009moments,anstreicher2012convex,pena2015completely} or matrix factorisation \citep{bach2008convex}.

\section{problem, reformulation, limiting convexity, \texorpdfstring{\&}{and} outline}

In this section we outline the version of dictionary learning we study and provide novel reformulations, one entirely in terms of the latent code and another that is convex for wide dictionaries.

%In this section we outline the dictionary learning problem, we show it has surprising convexity in the many neuron limit, and we describe our key results: necessary conditions on local optimality. 

\textbf{Dictionary Learning \& Notation} The basic assumption of both dictionary learning and the linear representation hypothesis is that data, $\vx^{[i]}\in\mathbb{R}^{d_x}$ - in this case, a neural network representation - can be well modelled as a linear combination of a set of sparse components: $\vx^{[i]} \approx \mW\vz^{[i]} + \vb$. Here, $\vz^{[i]}\in\mathbb{R}^{d_z}$ is the (hopefully) sparse SAE representation of datapoint $i$, $\mW$ is the weight matrix/dictionary with columns $\vw_d$ that approximately maps $\vz^{[i]}$ to $\vx^{[i]}$, $\vb$ is a bias vector, $d_x$ and $d_z$ are the size of the neural network representation and the SAE respectively (generally $d_z \gg d_x$), and $i$ runs from $1$ to $N$, the number of datapoints. Further, we will denote with $\vz_d\in\mathbb{R}^N$ the vector of responses of SAE neuron $d$. To learn this decomposition we optimise the following loss:
\begin{equation}\label{eq:dictionary_learning_problem}
    \min_{\mW, \{\vz^{[i]}\}_{i=1}^N, \vb} \mathbb{E}_i\bigg[||\mW\vz^{[i]} + \vb -\vx^{[i]}||_2^2 + 2\lambda||\vz^{[i]}||_1\bigg], \qquad \text{subject to} \quad \vz^{[i]} \geq 0, \quad ||\vw_d||_2 = 1
\end{equation}
The objective has two terms, the reconstruction error that ensures accurate fitting of the data, and an L1 penalty that encourages sparseness (we define $\lambda$ with a $2$ for later convenience). Without constraints, a trivial approach to decreasing the loss involves simultaneously scaling up $\mW$ and down $\vz^{[i]}$. Following \citet{bricken2023towards}, we control this by constraining each dictionary column, $\vw_d$, to be unit norm. Finally, $\vz^{[i]}\geq 0$ is a common but non-essential choice that reflects the idea that components should contribute in one direction. In the following, we will describe constraints on feature directions, $\vw_d$, and distributions, $\vz_d$, implied only by optimising \cref{eq:dictionary_learning_problem}.

\textbf{Encoder Architecture} While linearity from $\vz^{[i]}$ to $\vx^{[i]}$ reflects the key representational hypothesis, the encoder from $\vx^{[i]}$ to $\vz^{[i]}$ is less specified. Standard SAEs learn a separate affine+ReLU transformation: $\vz^{[i]} = \text{ReLU}(\mW_{\text{enc}}\vx^{[i]} + \vb_{\text{enc}})$ \citep{bricken2023towards,cunningham2023sparse,templeton2024scaling}, which additionally constrains the learnt concepts. Motivated partly by this gap between ideal dictionary learning representations, i.e. solutions to~\cref{eq:dictionary_learning_problem}, and those learnt by a standard SAE, various alternate encoders \citep{hindupur2025projecting}, losses \citep{fel2025archetypal}, or nonlinearities \citep{gao2024scaling,bussmann2024batchtopk,rajamanoharan2024jumping} have been considered, including sparse coding algorithms which can close this gap \citep{costa2025flat,o2024compute}.

Despite this architectural diversity, we study solutions to the vanilla dictionary learning problem,~\cref{eq:dictionary_learning_problem}, equivalent to encoders with no optimisation gap \citep{o2024compute}. We view this as an idealised theoretical target demonstrating how the bare bones of dictionary learning suffice to heavily structure optimal representations. That said, some of our results, e.g. the following reformulations, hold if we add the encoder constraint, $\vz^{[i]} = \text{ReLU}(\mW_{\text{enc}}\vx^{[i]} + \vb_{\text{enc}})$, and there are certain settings in which standard SAEs exactly solve \cref{eq:dictionary_learning_problem} (see \cref{app_sec:problem_statement}). 

\textbf{Loss Reformulation} In~\cref{app_sec:symmetry_reformulations}, we use a scaling symmetry argument from \citet{bach2008convex} to develop two equivalent formulations of the original problem,  \cref{eq:dictionary_learning_problem}. First, we show that, for the following problem studied by \citet{templeton2024scaling} in which the norm of the dictionary elements are regularised rather than constrained, the solutions are equivalent up to scaling:
\begin{equation}\label{eq:unconstrained_norm_dictionary_learning}
    \min_{\mW, \{\vz^{[i]}\}_{i=1}^N, \vb} \mathbb{E}_i\Big[||\mW\vz^{[i]} + \vb -\vx^{[i]}||_2^2\Big] + 2\lambda\sum_{d=1}^{d_z}||\vw_d||_2\mathbb{E}_i[z_d^{[i]}], \qquad \text{subject to} \quad \vz^{[i]} \geq 0
\end{equation}
Hence, while changing from~\cref{eq:dictionary_learning_problem} to~\cref{eq:unconstrained_norm_dictionary_learning} aids optimisation \citep{templeton2024scaling}, it does not change the optima, a point likely well-understood by the original authors, though not stated.

A second reformulation related by scaling uses the same trick to separate activity and weight norms:
\begin{equation}\label{eq:separate_regularisor_dictionary_learning}
    \min_{\mW, \{\vz^{[i]}\}_{i=1}^N, \vb} \mathbb{E}_i\Big[||\mW\vz^{[i]} + \vb -\vx^{[i]}||_2^2\Big] + \frac{\lambda}{N}\sum_{d=1}^{d_z}(||\vw_d||^2_2 + ||\vz_d||_1^2), \qquad \text{subject to} \quad \vz^{[i]} \geq 0
\end{equation}
This loss is an unconstrained quadratic in the dictionary, $\mW$, so we can optimise it out, \cref{app_sec:Z_reformulations}. Denoting with $\mZ\in\mathbb{R}^{d_z \times N}$ and $\mX\in\mathbb{R}^{d_x \times N}$ the matrices formed by stacking SAE and network representations respectively, and $\bar\mZ$ and $\bar\mX$ their row-demeaned versions, the optimal dictionary is $\mW^* = \bar\mX(\lambda\mathds{1}+\bar\mZ^T\bar\mZ)^{-1}\bar\mZ^T$, giving our third reformulation solely in terms of the latent code:
\begin{equation}\label{eq:only_Z_dictionary_learning}
    \min_{\mZ}\Tr[(\lambda\mathds{1}+\bar\mZ^T\bar\mZ)^{-1}\bar\mX^T\bar\mX] + \mathbf{1}^T\mZ^T\mZ\mathbf{1}, \quad   \text{subject to} \quad \mZ \geq 0
\end{equation}

\textbf{Convex Reformulation} In \cref{app_sec:Q_reformulations}, we note that the problem, \cref{eq:only_Z_dictionary_learning}, can be written entirely in terms of the representational similarity, $\mQ = \mZ^T\mZ\in\mathbb{R}^{N\times N}$, and its demeaned version $\bar\mQ$:
\begin{equation}\label{eq:only_Q_dictionary_learning}
    \min_{\mQ}\Tr[(\lambda\mathds{1}+\bar\mQ)^{-1}\bar\mX^T\bar\mX] + \mathbf{1}^T\mQ\mathbf{1}, \quad   \text{subject to} \quad \mQ \in \mathcal{CP}^N, \quad \text{cp-rank}(\mQ) \leq d_z
\end{equation}
$\mZ \geq 0$ has become the constraint that $\mQ = \mA^T\mA$ for some $\mA \geq 0$. Matrices with this property are called completely positive, denoted by $\mQ \in\mathcal{CP}^N$ \citep{berman2003completely} and form a convex set. Further, the objective is a convex function of $\mQ$ \citep{dorrell2026convex}. The remaining non-convexity is the restriction $\text{cp-rank}(\mQ) \leq d_z$ on the completely positive rank of $\mQ$, the minimal number of rows of $\mA\geq 0$ such that $\mQ = \mA^T\mA$. The cp-rank of an $N
\times N$ matrix is bounded by $N(N+1)/2$ \citep{shaked2013cp}, so as long as $d_z 
\geq N(N+1)/2$ we can drop this and the optimisation is convex, ensuring local minima are global. Hence, in the unrealistic very wide setting, the dictionary learning problem can be reformulated as convex. %Relative to the convex problems explored by \citet{bach2008convex}, nonnegativity frames a particularly clean convex problem using $\mathcal{CP}^N$, though 
While interesting, determining membership of $\mathcal{CP}^N$ is NP-Hard \citep{dickinson2014computational} making this more useful theoretically than practically. We will use this framing to study the wide-dictionary limit in \cref{sec:feature_splitting_limit}. 

\textbf{Vanilla SAEs} Finally, by adding the encoding constraint $\vz^{[i]} = \text{ReLU}(\mW_{\text{enc}}\vx^{[i]} + \vb_{\text{enc}})$, each of the above problems can be turned into its vanilla SAE analogue. Pleasingly, thanks to the homogeneity of ReLU, the preceding arguments continue to hold, including convexity over $\mQ$ \citep{dorrell2026convex}.

\textbf{Outline} Our analyses now uses these different reformulations to explore a series of phenomena:
\begin{itemize}
    \item In \cref{sec:feature_feature}, we present a particular first-order optimality condition constraining the relationships between optimal features. We use this to identify separable feature sets, explain hierarchical splitting and absorption, and broadly get a handle on SAE feature preferences.

     \item In \cref{sec:error_residuals} we present similar first-order optimality conditions relating features to residuals, constraining the relationship between what an optimal dictionary does and doesn't encode.

     \item In \cref{sec:dense_antipodal_feature_pairs} we turn to the finding of dense antipodal pairs of features, and explain this using a completely solvable toy model of rank 1 data.
 
     \item In \cref{sec:feature_splitting_limit} we study the wide convex limit of dictionary learning, and find that data is split into rays, reinforcing the width-dependence of discovered features.
\end{itemize}

\section{feature constraints, and hierarchical splitting/absorption}\label{sec:feature_feature}

Here, we describe how optimality forces features to inactivate separately, explaining feature splitting.

\begin{figure}[!b]
    \centering
    \includegraphics[width=\linewidth]{Figures/ICLR_FIG1.png}
    \caption{\textbf{A)} We plot two putatively optimal sparse independent features, normalised according to \cref{eq:feature_feature_constraint}. The relevant convex hulls (red regions) are the region explored by one feature while the other is inactive. In this case the convex hulls cover $[-1, 1]$, so these features are stable for all $\cos(\theta_{12})$. \textbf{B)} Features that never inactivate together are unstable when positively aligned (purple region), while \textbf{(C)} correlated features can be easily stabilised by points that expand the convex hull. \textbf{D)} We train a two element dictionary on the datasets in \textbf{A-C} with various alignments $\cos(\theta_{12})$. We compute an index that measures how amodular the representation is (\cref{sec_app:feature_experiments}, 0 = each latent has its own dictionary element). We see meaningful amodularity only when predicted from the convex hulls, \cref{sec_app:feature_experiments}. \textbf{E)} A categorical feature that is more than half sparse is stable independent of the encoding geometry. \textbf{F)} Hierarchical features are unstable under all cosine vectors except perfect alignment, whereas \textbf{(G)} feature absorption turns the hierarchy into mutually exclusive features, stabilising them.}
    \label{fig:feature_fig}
\end{figure}

\textbf{Local Optimality Conditions} If optimised correctly, a dictionary learning solution will converge to a local minima. Given just local optimality, what can we say about the representation, $\mZ$? It turns out, quite a lot. Being locally optimal means there is no perturbation that, while preserving nonnegativity, decreases the loss to first order. In \cref{sec_app:local_optimality_conditions} we study perturbations and, using either the subgradients of non-differentiable functions or dual feasible certificates, derive conditions on $\mZ$ such that no loss-decreasing perturbation exists, as in \citet{gribonval2010dictionary}. We present the general perturbation in \cref{sec_app:general_perturation}, which recovers the classic ISTA-like algorithm \citep{daubechies2004iterative} and gets the closest to a sufficient condition. Here, we consider a more specialised result that constrains the relationship between learnt features, and in~\cref{sec:error_residuals} we consider feature-residual relations.

\textbf{Necessary Feature Relations} In \cref{sec_app:feature_feature_perturbation}, we study a perturbation within the span of the features, $\mZ$. From this we derive the following necessary condition relating features to one another.
\begin{equation}\label{eq:feature_feature_constraint}
    -\hat\mW^T\hat\vw_d \equiv -\cos(\vtheta_d) \in \text{Convex Hull}\bigg(\bigg\{\frac{\bar\vz^{[i]}}{|\min_j \bar\vz^{[j]}|}\bigg| z^{[i]}_d = 0\bigg\}\bigg)
\end{equation}
where the $\min$ and division are taken elementwise and $\hat\vw_d$ denotes the unit-scaled $\vw_d$. Since \citet{gribonval2010dictionary} consider a complete, basis-forming dictionary, this is an exact analogue of their result in our slightly different problem setting. The left hand side is a $d_z$-dimensional vector of cosine similarities between $\vw_d$ and each of the other dictionary elements. The equation states that this vector of cosines must lie within the convex hull of the normalised data observed when feature $d$ is off. This enforces a relationship between optimal features' directions and joint-distributions. %\will{In figure blah we show, despite architectural misspecification, this relationship holds in a large-scale SAE.}

\textbf{Intuition} Consider a 2-element dictionary and with only one cosine similarity: $\cos(\theta_{12}) = \hat\vw_1^T\hat\vw_2$. The red regions in \cref{fig:feature_fig}A illustrate the relevant convex hulls for two independent sparse variables: the ranges that each variable explores while the other is inactive. In this case the convex hulls span the entire $[-1, 1]$ region, so every choice of $\cos(\theta_{12})$ is allowed under \cref{eq:feature_feature_constraint}, reflecting the fact that sparse, independent features are stably extracted by dictionary learning. However, it also motivates representations that won't be local optima. Consider two features that never inactivate together, \cref{fig:feature_fig}B; the convex hull will lose its minimal region, highlighted in purple, destabilising features encoded along highly aligned directions. Roughly, aligned encoding directions reflect the fact these features are explaining similar data, and since they don't inactivate together they can be merged while maintaining nonnegativity without incurring too much cost.
%\footnote{Intuitively, since they are explaining similar data they can be merged, and since they don't inactivate together, keeping the combination nonnegative won't incur too much activity cost.}
Alternatively, two features can be very highly correlated while still satisfying \cref{eq:feature_feature_constraint}, as long as they activate and inactivate together, \cref{fig:feature_fig}C. \cref{fig:feature_fig}D shows feature recovery when training a two element dictionary on these datasets for different values of $\cos(\theta_{12})$: the features are stable only when the convex hull condition is satisfied.

\textbf{Acceptable Features} Which kinds of features pass \cref{eq:feature_feature_constraint}'s conditions? The convex hull construction, e.g. \cref{fig:feature_fig}C, emphasises the role of data range. Indeed, a sufficient condition for satisfying \cref{eq:feature_feature_constraint} {\it for any set of encoding directions} is that while each feature is off the others explore the extremities of their range, as in \cref{fig:feature_fig}A \& C. This dataset property is a variant of range-independence, which drives feature recovery in other settings \citep{roth2023disentanglement,dorrell2025range}; here too we find dictionary learning stabilises features in which turning one element off leaves the range of the others unchanged. Alternatively, categorical features (`mammal', `fish', etc.) are stable when orthogonally embedded, and if they are all inactivate at least half the time (i.e. no more than half the tokens are animals) they are stable for all encoding geometries, \cref{app_sec:specific_data_distributions}, 
\cref{fig:feature_fig}E. Finally, in \cref{app_sec:specific_data_distributions} we show these conclusions generalise to separable data, a generalisation where categories are no longer mutually exclusive, but in which each feature activates alone in at least one datapoint \citep{donoho2003does}. In conclusion, range-independent or categorical variables can easily pass.

%Conversely, if the features are roughly orthogonal more typical feature relations satisfy \cref{eq:feature_feature_constraint}; for example, mutually exclusive categories of any dimensionality, or a dataset in which all features activate significantly in isolation, the familiar separability condition from the Nonnegative Matrix Factorisation literature \citep{donoho2003does}, \cref{app_sec:specific_data_distributions}. More generally, most existing identifiability results tackle this near-orthogonal regime (e.g. \citep{gribonval2010dictionary,wu2018local,wang2020unique}), so their results could perhaps be mapped onto this geometric picture. Finally, a set of mutually exclusive features that are inactive at least half the time are 

\textbf{Hierarchical Splitting \& Absorption} Two striking SAE observations are feature splitting, and its cousin, feature absorption. In feature splitting, a broad concept that appears in a narrow SAE, like `maths', is seen to split into more fine-grained concepts in larger SAEs, like `geometry' or `topology' \citep{bricken2023towards}. Feature absorption, on the other hand, describes how in a single SAE with coexisting overlapping high and low level features, such as `word starting with S' and the word `short', the high level feature unexpectedly won't co-activate with the lower \citep{chanin2024absorption}. Both of these fly in the face of intuition: feature absorption creates unintuitive gaps in the activity of the higher level feature (a `starting with S' feature that doesn't activate to the word `short'), while feature splitting ensures SAEs miss the wood for the trees \citep{leask2025sparse}. Indeed, there has been SAE development motivated by exactly these frustrations \citep{bussmann2025learning,costa2025flat}.

Existing work has outlined the pressures that lead to this phenomenon. \citet{till_2024_saes_true_features} argues that, with enough neurons, an SAE should encode feature combinations rather than the features themselves, while \citet{chanin2024absorption} show mathematically that if the low level feature activates only when the high level feature is active, feature absorption decreases the loss. Here, we present these as natural conclusions of optimal feature-feature conditions, \cref{eq:feature_feature_constraint}, relating and generalising the findings.

To see this, consider the joint distribution of two hierarchically related features, which we operationalise as a lower level feature, such as `labrador', that activates only when a higher level, such as `dog', is active \cref{fig:feature_fig}F. Since all labradors are dogs, the distribution of `labrador' while 'dog' is inactive, one of the convex hulls appearing in \cref{eq:feature_feature_constraint}, is a single point where both features are inactive. This ensures that, unless they align completely making them identical, no matter their cosine similarity, hierarchically related features are unstable and will not appear in optimal representations! %\footnote{Strictly speaking, since the convex hull contains one point, there could still be one stable cosine similarity. But this is a meaningless case corresponding to completely aligning the vectors, making `labrador' and `dog' identical.}! 

Feature splitting stabilises hierarchically related features by constructing a feature for each lower-level feature. This makes the features mutually exclusive \cref{fig:feature_fig}E, exactly those that dictionary learning naturally stabilises. 
Finally, feature absorption can be seen as a similar route to stability: by absorbing the instances of co-activity into the lower-level feature, the representation has again been turned into a set of mutually exclusive categories, \cref{fig:feature_fig}G. Feature absorption is simply feature splitting but using a lower-level category corresponding to `not one of the other low-level categories'.

\begin{figure}[!b]
    \centering
    \includegraphics[width=0.9\linewidth]{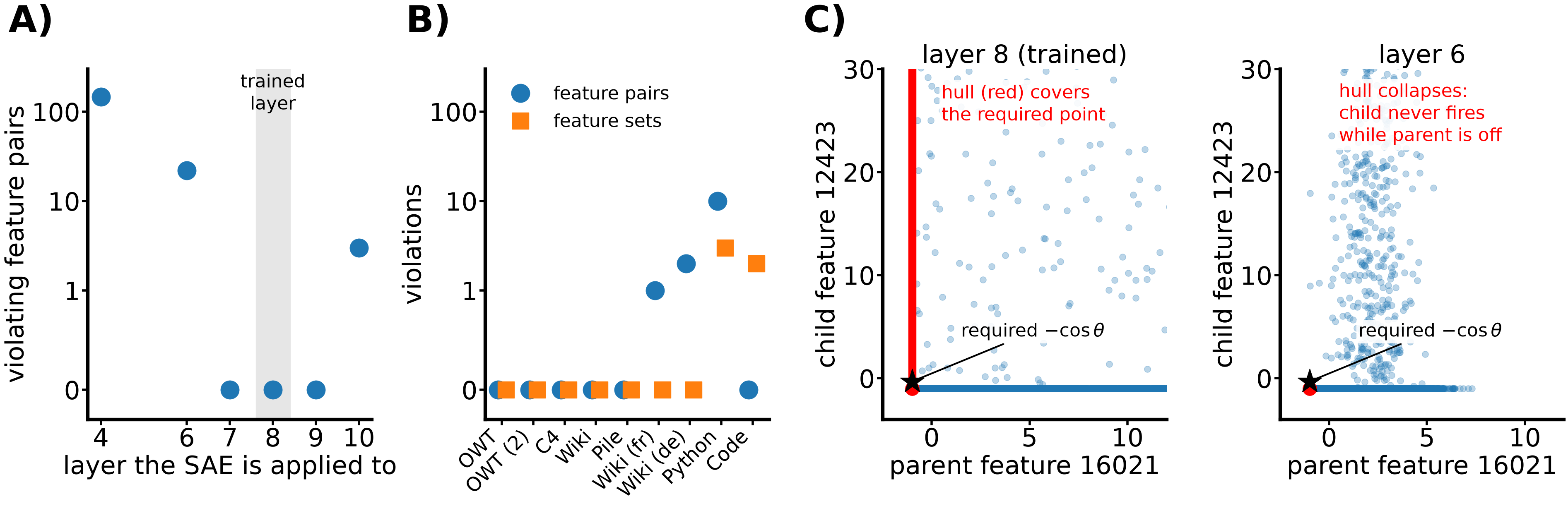}
    \caption{\textbf{A)} We apply an SAE trained on layer 8 to other layers and find the number of pairs of features that break \cref{eq:feature_feature_constraint} increases with distance from layer 8. Similarly, \textbf{(B)} when applying an SAE trained on OWT to other datasets the number of violating feature pairs (blue dots) or sets of features (orange squares, we did not test exhaustively) increases with semantic distance from OWT. \textbf{C)} An example feature pair that within distribution satisfies the convex hull conditions, no longer does when tested out-of-distribution, as the child never fires while the parent is inactive.}
    \label{fig:SAE_data_fig}
\end{figure}

\textbf{Real SAEs} We tested whether \cref{eq:feature_feature_constraint} is satisfied for vanilla SAEs trained on block 8 of GPT2-small's residual stream using OpenWebText (OWT) \citep{bloom2024gpt2residualsaes,bloom2024saetrainingcodebase}, and found no pairwise violations among 3700 sufficiently active features (14M pairs) nor in sets of up to 100 sampled features, \cref{app_sec:real_SAE_experiments}. To show that these conditions are not vacuous, we apply the layer-8 SAE off-distribution, either to other layers, or to other datasets. As the test layer moves further from train the number of pairwise violations increases, \cref{fig:SAE_data_fig}A. Similarly, while roughly similar datasets like English wikipedia show no pairwise violations, French or German wikipedia, and code datasets do, even when testing only the pairwise version of \cref{eq:feature_feature_constraint}, \cref{fig:SAE_data_fig}B. These violations tend to be hierarchical, like a sparse `french loan word feature' that activates widely on french wikipedia including at all times when a grammatical feature, like ``as/like'' activates. The analogous code example is a parent encoding stray space in prose or indentations in code, making it a strict superset of another feature that fires within nested code indentations (though on OWT it encodes islands and beaches...).

\section{Error Residuals}\label{sec:error_residuals}

Next, we consider first-order conditions on the feature-residual relationships, these preclude the dictionary from learning concepts that are too similar to patterns it leaves in the residuals.

\textbf{Constraining Feature-Residual Relationships} In \cref{sec_app:feature_residual_perturbation} we show that the general first-order stability condition, in which we consider all perturbations to $\mZ$ that preserve nonnegativity, can be rewritten as a constraint on the behaviour of residuals:
\begin{equation}\label{eq:feature_residual_constraint}
    -\frac{1}{N}\hat{\vw}_d^T(\bar\mX - \mW\bar\mZ)\delta \vz = -\frac{1}{N}\hat{\vw}_d^T\mE\delta \vz \in \lambda \text{Convex Hull}(\{\delta \vz_i| z^{[i]}_d = 0\})
\end{equation}
where $\delta \vz$ is an arbitrary mean-zero response vector and we've defined the residual matrix, $\mE$. This shows a natural dependence on $\lambda$: harsher regularisation means looser constraints on the residuals.

\textbf{Intuition} $\delta \vz$ is a mean-zero pattern of activity that could be added to neuron $d$'s response. Doing so would change both the reconstruction and regularisor. $\hat{\vw}_d^T\mE$ is the residual along neuron $d$'s direction, hence $\frac{1}{N}\hat{\vw}_d^T\mE\delta \vz$ is the correlation between the projected error and the new activity $\delta \vz$, which determines how much better the reconstruction would be if this perturbation were added. Conversely, adding $\delta \vz$ to $\vz$ while preserving non-negativity requires shifting by a bias, which incurs an additional L1 cost whose size is determined by set of values $\delta \vz$ takes while $\vz_d$ is inactive. If the gain from reconstruction is smaller than the L1 cost the left hand side will lie within the convex hull on the right hand side, the condition will be satisfied, and the representation will be stable. 

\begin{figure}[!bh]
    \centering
    \includegraphics[width=0.95\linewidth]{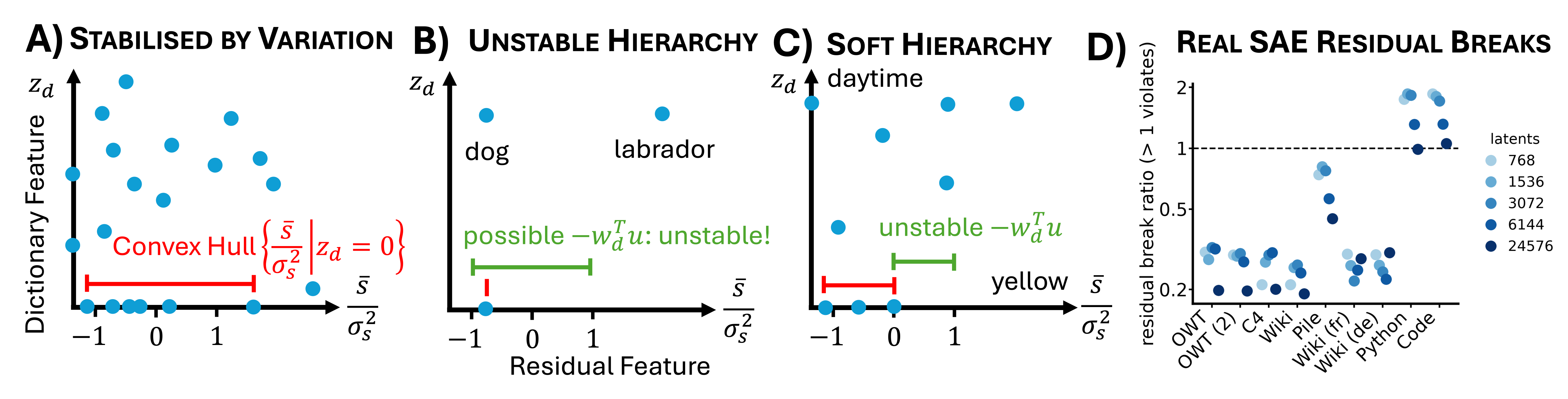}
    \caption{\textbf{A)} We plot a dictionary against missing feature; if the missing feature varies widely enough while the dictionary is inactive, it can be stably left in the residuals  \cref{eq:feature_residual_constraint}. \textbf{B)} Hierarchical features don't survive this check, while \textbf{(C)} soft hierarchies fail in one direction. \textbf{D)} For OWT-trained SAEs across sizes we find residual breaks in code datasets.}
    \label{fig:residual_fig}
\end{figure}

Let's break the residual into singular vectors, $\mE = \sum_i \vu_i\bar\vs_i^T$ ($||\vu|| = 1$ without loss of generality), and, considering each $\bar\vs_i$ as a missing feature, derive an interesting necessary condition using $\delta\vz = \bar\vs$:
\begin{equation}\label{eq:residual_constraint}
    - \hat\vw_d^T\vu \in \lambda \text{Convex Hull}\bigg(\bigg\{\frac{1}{\sigma_s^2 } \bar s^{[i]}| z^{[i]}_d = 0\bigg\}\bigg), \quad \sigma_s^2 = \frac{1}{N}||\bar\vs||^2
\end{equation}
Thus, to be stable, the cosine similarity between the missing concept's direction and $\vw_d$ must be within the range $\bar\vs$ takes while the neuron is inactive scaled by $\frac{\lambda}{\sigma_s^2}$. %If, on each neurons' inactive set $\vs$ varies widely this dictionary is likely stable. If $\bar\vs$ is only takes certain values it is likely unstable. 
We present three implications:
\begin{itemize}
    \item Similar to our earlier range-independent results, if a feature varies widely while each dictionary element is off, it can be stably left in the residual, \cref{fig:residual_fig}A.% and this margin for error widens as the regularisation strengthens (larger $\lambda$) or the variance shrinks (smaller $\sigma_s^2$).
    \item Range-dependency destabilises the solution. Hierarchical features are again problematic, if `labrador' is off whenever `dog' is, an optimal dictionary containing a `dog' feature is forced to mop the `labrador' direction into the `dog' feature, no matter how orthogonal, \cref{fig:residual_fig}B.
    \item Softer hierarchies are destabilised one way. If a sub-feature (`yellow') only activates maximally with a super-feature (`daytime'), otherwise activating weakly then a `daytime' neuron would be destabilised by the `yellow' residual, but only if they anti-align, \cref{fig:residual_fig}C.
\end{itemize}
We again test \cref{eq:feature_residual_constraint} on the OWT-trained SAE tested on other datasets by measuring a residual break measure of violation, \cref{app_sec:real_SAE_experiments}. We found that the clearest violations again arose from features that activated to whitespace in code datasets, but formatting or byte-level information in OWT, reflecting missing representational structure could usefully have been captured in the code datasets.

\section{Dense Antipodal Feature Pairs}\label{sec:dense_antipodal_feature_pairs}

Dense features provide a case study in SAEs behaving strangely when applied to datasets which don't match sparsity assumptions, which we show can be understood simply as a consequence of dictionary learning optimality. \citet{sun2025dense} find that SAEs trained on the residual stream of Gemma 2 \citep{team2024gemma} learn dense latents that activate on 10 to 50\% of datapoints. Further, when using a nonnegative activation function, like JumpReLU or TopK, these latents form antipodal pairs whose dictionary vectors, $\vw_d$, point in opposite directions. But these pairs disappear when using an activation function that permits both positive and negative latents, such as AbsoluteTopK which thresholds features using the absolute value, rather than a ReLU \citep{sun2025dense}.

\begin{figure}[th]
    \centering
    \includegraphics[width=\linewidth]{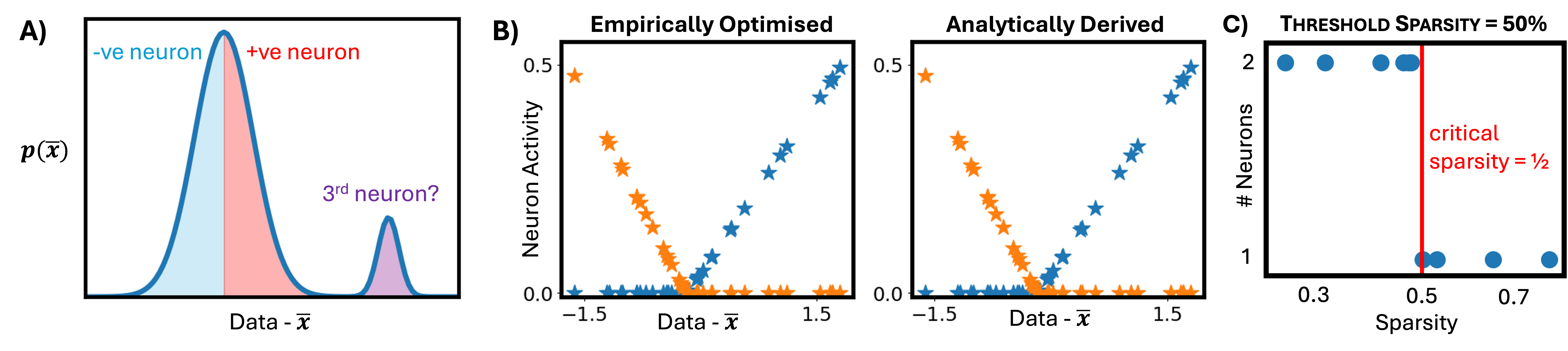}
    \caption{\textbf{A)} While splitting a dense variable into positive and negative halves seems intuitive, perhaps outlying clumps should have their own neuron? \textbf{(B)} We find this is not the case, an optimal solution to \cref{eq:only_Z_dictionary_learning} for rank 1 data exists with at most 2 non-zero neurons. This theoretically predicted low-neuron representation, \cref{sec_app:dense_encodings}, fits exactly. \textbf{C)} Theory predicts that a variable that is more than half sparse should be encoded with only one active neuron: empirics agree.}
    \label{fig:dense_antipodal}
\end{figure}

This result is somewhat intuitive: if forced to code a dense variable into a nonnegative sparse code, one might try breaking the variable into a positive and negative part, something made unnecessary by using AbsoluteTopK. We wondered whether this would always be the case? Intuitively, it seems non-obvious that two antipodal features are sufficient - if there is a small cluster of very high activity along one feature dimension perhaps the cost can be reduced by encoding it separately, \cref{fig:dense_antipodal}A?

We find this is not the case. In \cref{sec_app:dense_encodings} we study a simple example of a vanilla SAE encoding a single dimension, a scalar datapoint $x^{[i]}$. The sparsity of this variable is defined as the proportion of datapoints for which it equals its minimal value. In this case we can analytically solve the KKT conditions for the optimal SAE, which contains either one neuron if $x^{[i]}$ is sparse, or two when $x^{[i]}$ is dense, and these two neurons have antipodal decoder directions matching both empirics \citep{sun2025dense} and theory on related neural coding problems \citep{karklin2011efficient,barrett2016optimal,dorrell2026convex}, \cref{fig:dense_antipodal}B. In the small $\lambda$ limit, we derive the threshold between the one and two neuron regimes: a variable that is active less than half the time will be given one neuron, more and it will be broken into pairs, something we confirm in simulation, \cref{fig:dense_antipodal}C. This matches the finding that the dense features never activate more than 50\% of the time \citep{sun2025dense}. In the following section, we understand this as a special case of a broader phenomenon: ray-clustering.
%This result cannot be taken entirely for granted. For example, let’s say you cap the magnitude of feature activities, as in a biological neuron. This leads to an SAE with a saturating nonlinearity which uses more than two neurons to encode a dense feature. Dense antipodal pairs therefore arise from the interplay of linearity and nonnegativity-alone, an intuitive conclusion.

%and the analysis points to various ways to break it. First, the simple nonnegativity constraint is key; instead, we consider capping the magnitude of feature activities, as in a real biological neuron. This leads to an SAE with a saturating nonlinearity which sometimes uses more than two neurons to encode a single feature \will{FIG}. Second, it reflects how linearity constrains magnitude based decompositions, like that proposed in \will{FIG}. This is not chosen because neuron 3 would have to activate very strongly to produce such a large output. If, however, we provide each feature with a constant \cref{sec_app:dense_encodings_saturating_nonlinearity}. Dense antipodal pairs are therefore linked to using only a nonnegativity constraint $\vz \geq 0 $, an intuitive conclusion.}  

\section{ray-clustering and the wide convex limit}\label{sec:feature_splitting_limit}

\begin{figure}[!tb]
    \centering
    \includegraphics[width=\linewidth]{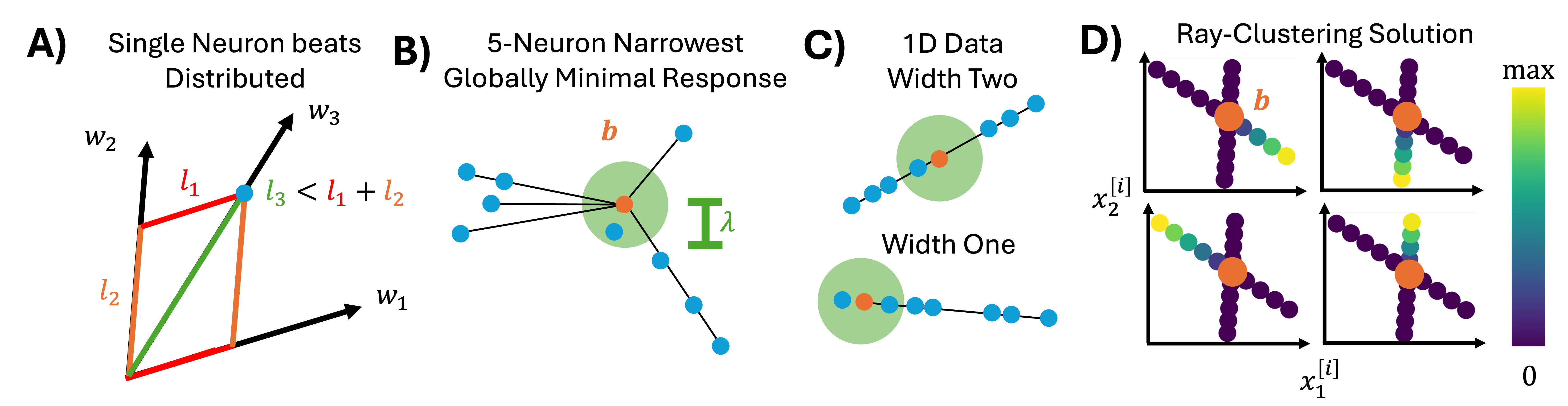}
    \caption{\textbf{A)} If a datapoint activates two neurons, and neurons are in abundance, it can be equally well-explained by a third that is aligned and hence uses less activity. \textbf{B)} If datapoints share rays from the optimal bias, $\vb$, it can be explained using only one neuron, narrowing the threshold number of neurons at which the dictionary transitions to the wide globally-minimal solution. Further narrowing comes from data within a $\lambda$-ball of $\vb$ that doesn't have to be encoded at all. \textbf{C)} In 1D this becomes the one or two neuron solutions from \cref{fig:dense_antipodal}. \textbf{(D)} We train a 30 neuron dictionary on this 2D cross dataset. We find only the four pictured neurons activate. The geometric median lands in the middle of the cross, and an optimal dictionary uses only four active neurons to explain the four rays of the data.}
    \label{fig:limiting}
\end{figure}

Finally, we show how dictionary learning optimality can be used to reason about the progressively finer feature splitting observed in SAEs. This feature splitting raises a question: as you increase dictionary size, will splitting continue {\it ad infinitum}? In short, yes. In the wide convex limit there exists a globally optimal one-dictionary-atom-per-datapoint solution \citep{bach2008convex}. For a simple route to this conclusion, consider the original objective, \cref{eq:dictionary_learning_problem} and imagine a solution in which two distinct dictionary atoms (i.e. with different dictionary vectors $\vw_d$) activate for a single datapoint, \cref{fig:limiting}A. This solution's L1 cost can be improved by instead fitting that datapoint with the same accuracy using a single dictionary element aligned with the datapoint, \cref{fig:limiting}A, leading to  the one-latent-per-datapoint.
%Hence, at a global minima each datapoint is encoded only by latents whose dictionary vectors align with the datapoint.
%In \cref{app_sec:more_neurons_than_datapoints} we follow a more convoluted argument the form of the optimal one-neuron-per-datapoint solution, and show that it is indeed a local, and from convexity therefore global, minima. 

However, \cref{sec:dense_antipodal_feature_pairs} teaches us that globally minimal solutions can sometimes be reached using much narrower dictionaries, down to 1 neuron. In general what governs how the minimal width globally minimal solution? We show it depends on three things: the regularisation hyperparameter $\lambda$, the optimal bias, $\vb$, and the number of rays from $\vb$ required to hit every datapoint further than $\lambda$ from $\vb$, \cref{fig:limiting}B. In \cref{app_sec:more_neurons_than_datapoints}, we derive $\vb$. Under perfect reconstruction ($\lambda\rightarrow 0$) it is the geometric median, the multidimensional generalisation of the median:
\begin{equation}\label{eq:geometric_median}
    \vb = \arg\min_\vb\sum_i||\bar\vx^{[i]} - \vb||_2
\end{equation}
Under imperfect reconstruction $\vb$ is a slight generalisation, \cref{app_sec:more_neurons_than_datapoints}. When $\lambda = 0$ the number of rays required to hit every datapoint from $\vb$ then determines the width of the narrowest global minima. This explains \cref{sec:dense_antipodal_feature_pairs}: in one dimension $\vb$ is simply the median; if more than 50\% of the data lies on an extreme point the median lies on an extreme point and only one ray is required, else, two are, \cref{fig:limiting}C.

At finite $\lambda$ we derive a second-order condition on whether an additional neuron can reduce the loss: 
\begin{equation}
    \label{eq:residual_wake_up_condition}
    \max_i ||\bar\vx^{[i]} - \mW\bar\vz^{[i]}||_2 \leq \lambda
\end{equation}
i.e. a solution is stable to new neurons if the residuals for all datapoints are below $\lambda$. If a locally optimal representation passes this test, it is a globally optimal solution, since it can be padded to with sufficient dead neurons to make the problem convex. This introduces a $\lambda$-buffer region of data around $\vb$ that don't activate any neurons, reducing the width of the narrowest global minima still further.

%Hence, having numerically solved this equation for $\vu$, the number of rays required to hit every datapoint determines the narrowest globally optimal solution, \cref{fig:limiting}B.

%In this section we consider arbitrarily large SAEs. We find that, in the limit of perfect reconstruction, a neuron-per-datapoint solution is globally optimal. However, we find two subtleties: first, as in \cref{fig:dense_antipodal}, sometimes the global optima can be reached with a much narrower SAE. Second, imperfect reconstruction clusters points together. This leads to the concept of the minimally-narrow globally-minimal SAE, which clusters data into 'ray-tubes', \cref{fig:limiting}C.

%\textbf{Neuron-per-Datapoint is Optimal} We'll consider the . Neurons are distinguished by $\vw_d$, and the loss is unchanged by shuffling activity within neurons that share a $\vw_d$ while preserving its sum. To focus on interesting behaviours, let's therefore assume each neuron has a distinct $\vw_d$. It is then clear that any solution in which there is more than one neuron active per datapoint can be improved by instead learning a feature specific to that datapoint, \cref{fig:limiting}A. 

\section{Discussion}

In this paper we studied how optimality in a dictionary learning problem, \cref{eq:dictionary_learning_problem}, structures learnt dictionaries, and we used this to puzzle over representational properties of SAEs - including hierarchical splitting \& absorption, the structure of residuals, dense antipodal features, and the neuron-per-datapoint-limit. This logic is pleasingly simple, unifying diverse behaviours in a single framework. 

Our results have clear limits. The conditions are necessary rather than sufficient, are derived for vanilla dictionary learning rather than specific encoders, and are tested only pairwise and on subsets of features. Two directions seem particularly useful. First, applying similar approaches to encoder architecture could link the choice of SAE to its representational assumptions, as we have done here for dictionary learning. Second, it remains unclear whether the linear representation hypothesis and feature superposition accurately reflect how neural networks represent information. Tools that separate what is imposed by the SAE from what is present in the network are a prerequisite for answering this.

Most broadly, our observations are filtered through the tools we use and the assumptions they necessarily carry. Dictionary learning is at least clear about its assumptions, and this paper uses these assumptions to interpret observed phenomena as products of this filtering. Practically, the same conditions offer a cheap diagnostic for when an SAE's dictionary no longer fits the data it is applied to. More ambitiously, by emphasising the outsized role these assumptions play, we hope to motivate the development of alternative interpretability tools with more realistic representational assumptions. 

\newpage

\subsection*{AI use statement}

In this work we used generative AI tools for checking mathematical arguments, helping discover relevant papers and surveying the literature, poking holes in the paper, and coding up the experimental work. While AI has checked every line of the maths, it only wrote those that appear in \cref{app_sec:specific_data_distributions}. The authors are to blame for most of the logic barring one very helpful case, where Claude recognised the geometric median and provided links to its speedy computation, a concept we had never heard of before. Finally, we did not write the code that was used to generate the non-toy experiments. Needless to say, we take responsibility for the final content of this work.

\subsection*{Reproducibility} 
All mathematical arguments are included in the appendix. Code to run experiments can be found in the following git repo: \url{https://github.com/WilburDoz/SAE_Identifiability.git}%anonymised git repo: \url{https://anonymous.4open.science/r/SAE_Identifiability-3243/README.md}.

\subsection*{Acknowledgements} 

I'd like to thank a large crowd of people for talking with me about this paper, especially for their guidance in exploring areas that were new to me: Priscila de Azevedo Drummond, James Whittington, Demba Ba, Shubham Choudhary, Sumedh Hindupur, Kyle Hsu, Luke Hollingsworth, Hadas Orgad, Bahareh Tolooshams, Valérie Costa and Charlie O'Neill. Further, special thanks go to Rodrigo Carrasco-Davis and Andy Arditi for comments on a draft of this paper, to Jin Hwa Lee for discussions of relevant SAE literature, and especially to Basile Confavereux for being the perfect partner-in-crime, were it not for all the very important other things he has to do ;)

I acknowledge funding from the Kempner Institute at Harvard University.

\bibliography{Support_Docs/bibliography}
\bibliographystyle{iclr2027_conference}

\newpage

\appendix

\section{Dictionary Learning Reformulations \texorpdfstring{\&}{and} Convexity}\label{app:loss_reformulations}

\subsection{Problem Statement}\label{app_sec:problem_statement}

We consider a standard formulation of dictionary learning. Given each of $N$ datapoints, $\vx^{[i]}\in\mathbb{R}^{d_x}$, we seek a sparse representation, $\vz^{[i]}\in\mathbb{R}^{d_z}$, from which we can linearly reconstruct the data:
\begin{equation}\label{eq_app:data_fitting_dict}
    \vx^{[i]} \approx \mW\vz^{[i]} + \vb
\end{equation}
where $\vb$ is a bias and $\mW$ is the dictionary matrix. For terminology, we will refer to each element of $\vz^{[i]}$ as a neuron or feature, and each column of the dictionary matrix, $\vw_d$, as the feature's direction or dictionary element. To enforce the dual goals of accurate fitting of data,~\cref{eq_app:data_fitting_dict}, and sparsity, we use the following optimisation problem,~\cref{eq:dictionary_learning_problem} in the main text:
\begin{equation}\label{eq_app:dictionary_learning_problem}
    \min_{\mW, \{\vz^{[i]}\}_{i=1}^N, \vb} \mathbb{E}_i\bigg[||\mW\vz^{[i]} + \vb -\vx^{[i]}||_2^2 + 2\lambda||\vz^{[i]}||_1\bigg], \qquad \text{subject to} \quad \vz^{[i]} \geq 0, \quad ||\vw_d||_2 = 1
\end{equation}
We make the standard substitution of enforcing sparsity using the continuous (and convex) L1 norm, rather than L0, for mathematical practicality, and we define $\lambda$ with a factor of $2$ for later convenience. The unit-norm constraint on the dictionary columns is introduced to circumvent the scaling degeneracy between representations, $\vz^{[i]}$, and dictionary elements, $\mW$. Finally, the non-negativity of the codes reflects a common intuition that each dictionary element should contribute only positively to the representation.

\citet{bricken2023towards} study exactly this optimisation problem with one important difference. Rather than inferring $\vz^{[i]}$ by solving~\cref{eq_app:dictionary_learning_problem}, they `amortise the inference' of $\vz^{[i]}$ via an encoder from $\vx^{[i]}$:
\begin{equation}\label{eq_app:encoder_SAE}
    \vz^{[i]} = \text{ReLU}(\mW_{\text{enc}}\vx^{[i]} + \vb_{\text{enc}})
\end{equation}
This combination,~\cref{eq_app:dictionary_learning_problem} and~\cref{eq_app:encoder_SAE}, represents the `vanilla sparse autoencoder' (SAE) problem. 

The Affine+ReLU encoder structure is important, placing constraints on the learnable SAE representations. However, we will ignore it, instead studying solutions to the original dictionary learning problem \cref{eq_app:dictionary_learning_problem}. In some cases this idealised model and vanilla SAEs are identical, for example, when the dictionary elements are orthogonal \cref{sec_app:general_perturation}, but this is not generally true. Instead this choice reflects an idealisation. There's a zoo of possible SAE architectures and the way architecture constrains learnt features has been studied\citep{hindupur2025projecting}. Instead, we ask how the dictionary learning objective alone shapes optimal representations. This is not an unfeasible idealisation, indeed algorithms exist to recover these optimal solutions, requiring only the addition of recurrent weights to the standard Affine+ReLU SAE \cref{app_sec:LCA}, and these have been used within the standard SAE pipeline \citep{o2024compute}. We therefore view vanilla dictionary learning as a good `model organism' to help us explain representational patterns across SAE architectures. 

\subsection{Reformulations Using Scaling Symmetry}\label{app_sec:symmetry_reformulations}

In this section we will use scaling symmetry to equate three versions of dictionary learning.

Consider the following optimisation problem,~\cref{eq:separate_regularisor_dictionary_learning} in the main text, that initially appears quite different to the original dictionary learning problem~\cref{eq_app:dictionary_learning_problem}:
\begin{equation}\label{eq_app:separate_regularisor_dictionary_learning}
    \min_{\mW, \{\vz^{[i]}\}_{i=1}^N, \vb} \mathbb{E}_i\Big[||\mW\vz^{[i]} + \vb -\vx^{[i]}||_2^2\Big] + \frac{\lambda}{N}\sum_{d=1}^{d_z}(||\vw_d||^2_2 + ||\vz_d||_1^2), \qquad \text{subject to} \quad \vz^{[i]} \geq 0
\end{equation}
We develop it using an argument from \citet{bach2008convex}. The reconstruction error is invariant to rescaling each neuron's activities and weights with a neuron-dependent scalar:
\begin{equation}
    \{\vw_d, \vz_d\}_{d=1}^{d_z} \rightarrow \{\sigma_d \vw_d, \frac{1}{\sigma_d}\vz_d\}_{d=1}^{d_z}
\end{equation}
That means we can optimise each $\sigma_d$ to minimise only the regularisation term. Performing this optimisation, one finds the optimal value is:
\begin{equation}
    (\sigma_d^*)^2 = \frac{||\vz_d||_1}{||\vw_d||_2}
\end{equation}
Inserting this value into~\cref{eq_app:separate_regularisor_dictionary_learning} we recover \cref{eq:unconstrained_norm_dictionary_learning} from the main text, the dictionary learning problem with unconstrained norm:
\begin{equation}\label{eq_app:unconstrained_norm_dictionary_learning}
    \min_{\mW, \{\vz^{[i]}\}_{i=1}^N, \vb} \mathbb{E}_i\Big[||\mW\vz^{[i]} + \vb -\vx^{[i]}||_2^2\Big] + \frac{2\lambda}{N}\sum_{d=1}^{d_z}||\vw_d||_2||\vz_d||_1, \qquad \text{subject to} \quad \vz^{[i]} \geq 0
\end{equation}
So, since any optimal solution will eventually settle to the optimal scaling $\sigma^*_d$, solving the original problem, \cref{eq_app:separate_regularisor_dictionary_learning}, is equivalent to solving \cref{eq_app:unconstrained_norm_dictionary_learning} then rescaling optimally. Finally, however, \cref{eq_app:unconstrained_norm_dictionary_learning} is now completely invariant to neuron-wise rescalings, so we can choose, for example, to only consider solutions in which $||\vw_d||_2 = 1$, recovering the original formulation,~\cref{eq_app:dictionary_learning_problem}. One thing this illustrates is that the SAE improvements reported by \citet{templeton2024scaling} upon moving from \cref{eq_app:dictionary_learning_problem} to \cref{eq_app:unconstrained_norm_dictionary_learning} must reflect better optimisations, since the properties of the optimal solutions are not meaningfully changed.

\subsection{Reformulation by Optimising over Decoder}\label{app_sec:Z_reformulations}

Now we will optimise over the decoder to define a problem only in terms of the sparse representation, again following the same path as \citet{bach2008convex}.

First, we introduce a few further matrices, $\mZ\in\mathbb{R}^{d_z\times N}$ and $\mX\in\mathbb{R}^{d_x\times N}$, the stacked SAE and network representation matrices, and their row-mean subtracted partners, $\bar\mZ$ and $\bar\mX$: \begin{equation}
    \bar\mZ = \mZ - \frac{1}{N}\mZ{\bf 1 1}^T
\end{equation}
We can optimise \cref{eq_app:separate_regularisor_dictionary_learning} directly over the bias $\vb$. Since it is unconstrained, it can mop up any errors in reconstruction along the mean direction, letting us write the optimisation problem as:
\begin{equation}
    \min_{\mZ, \mW} \frac{1}{N}||\mW\bar\mZ - \bar\mX||_F^2 + \frac{\lambda}{N}\big( \sum_{d=1}^{d_z}||\vz_d||^2_1 + ||\vw_d||_2^2\big), \quad \text{subject to} \quad \mZ \geq 0
\end{equation}
We see that we can now optimise directly over $\mW$, finding:
\begin{equation}\label{eq_app:optimal_dictionary_eqn}
    \mW^* = \bar\mX\bar\mZ^T(\lambda\mathds{1} + \bar\mZ\bar\mZ^T)^{-1} = \bar\mX(\lambda\mathds{1} + \bar\mZ^T\bar\mZ)^{-1} \bar\mZ^T
\end{equation}
Using this and dropping a shared $\frac{\lambda}{N}$ factor we find the following optimisation problem in $\mZ$:
\begin{equation}\label{eq_app:only_Z_dictionary_learning}
    \min_{\mZ} \Tr[(\lambda \mathds{1} + \bar\mZ^T\bar\mZ)^{-1}\bar\mX^T\bar\mX] + {\bf 1}^T\mZ^T\mZ{\bf 1}, \quad \text{subject to} \quad \mZ \geq 0
\end{equation}
Where we have rewritten the L1 loss penalty into a convenient matrix form. This is~\cref{eq:only_Z_dictionary_learning} from the main text, which we will use to derive our key results.

\subsection{Convexity in the Large Dictionary Limit}\label{app_sec:Q_reformulations}

Here we show that we can rewrite~\cref{eq_app:only_Z_dictionary_learning} in terms of the representational similarity, defined as:
\begin{equation}
    Q_{i,j} = \vz^{[i],T}\vz^{[j]}
\end{equation}
Surprisingly, and following existing work in dictionary learning \citep{bach2008convex} and theoretical neuroscience \citep{sengupta2018manifold,dorrell2026convex}, when we do this we find that the problem is convex when there are enough neurons. This is not a practical setting, but will prove useful to us in clarifying the structure of the loss landscape, and deriving limiting behaviours of SAEs.

\textbf{Activity Loss:} Using the non-negativity of $\vz_d$, we can rewrite the activity loss:
\begin{equation}
    \sum_{d=1}^{d_z}||\vz_d||_1^2 = \sum_{d=1}^{d_z}(\sum_{i=1}^Nz_{d}^{[i]})^2 =  \sum_{d,i,j=1}^{d_z,N,N}z^{[i]}_{d}z^{[j]}_{d}=  \sum_{i,j=1}^{N,N}(\vz^{[i]})^T\vz^{[j]} = \sum_{i,j=1}^{N,N}Q_{ij} = {\bf 1}^T\mQ{\bf 1}
\end{equation}
which is a linear, and therefore convex, function of $\mQ$.

\textbf{Nonnegativity Constraint} The set of dot product matrices with unconstrained ranks formed from vectors that are themselves nonnegative ($\mQ = \mZ^T\mZ$ for $\mZ \geq 0$) form a convex set called the set of completely positive matrices \citep{berman2003completely}. However, $\mQ$ must additionally be constrained to have a completely-positive rank lower than $d_z$, meaning that there exists a nonnegative matrix $\mA\geq 0$ with at most $d_z$ rows such that $\mA^T\mA = \mQ$. This is a non-convex constraint. The cp-rank of an $N\times N$ matrix is at most $N(N+1)/2$ \citep{shaked2013cp}, so when $d_z\geq N(N+1)/2$ we can drop the cp-rank constraint making the whole problem convex.

\textbf{Weight and Reconstruction Loss} The weight and reconstruction loss can be written in terms of $\mQ$:
\begin{equation}
    \min_{\mZ} \Tr[(\lambda \mathds{1} + \bar\mZ^T\bar\mZ)^{-1}\bar\mX^T\bar\mX] = \min_{\mQ} \Tr[(\lambda \mathds{1} + \bar\mQ)^{-1}\bar\mX^T\bar\mX]
\end{equation}
For completeness we prove convexity. Following \citet{dorrell2026convex}, we modify the proof technique of maths stack exchange user Robert Israel \citep{isreal2013trace}. A function, $\mathcal{L}$, of positive semi-definite matrix, $\bar\mQ$ is convex if for $\bar\mQ(t) = \bar\mQ+t\mS$ where $\mS$ is a symmetric matrix:
\begin{equation}
    \frac{d^2\mathcal{L}(\bar\mQ(t))}{dt^2}\bigg\vert_{t = 0} \geq 0
\end{equation}
Using the fact that:
\begin{equation}
    (\mA+t\mS)^{-1} = \mA^{-1} - t\mA^{-1}\mS\mA^{-1}+t^2\mA^{-1}\mS\mA^{-1}\mS\mA^{-1}
\end{equation}
with the identification $\mA = \lambda\mathds{1} + \bar\mQ$ we find:
\begin{equation}
    \frac{1}{2}\frac{d^2\mathcal{L}(\bar\mQ(t))}{dt^2}\bigg\vert_{t = 0} = \Tr[\mA^{-1}\mS\mA^{-1}\mS\mA^{-1}\bar\mX^T\bar\mX]
\end{equation}
Finally, writing $\mC = \mS\mA^{-1}\bar\mX^T$ this is:
\begin{equation}
    \frac{1}{2}\frac{d^2\mathcal{L}(\bar\mQ(t))}{dt^2}\bigg\vert_{t = 0} = \Tr[\mC^T\mA^{-1}\mC]
\end{equation}
But we know $\mA^{-1} = (\lambda\mathds{1}+\bar\mQ)^{-1}$ is positive definite, hence this quantity is nonnegative and the loss is convex.

Finally, this is a function on demeaned representational similarity matrices. Changing the mean changes $\mQ$, but doesn't change the weight loss, so as a function of the loss is, first, a demeaning projection (it is easy to check this is a projection), then the loss: 
\begin{equation}
    P: \mQ \rightarrow \mQ - \frac{1}{N}{\bf 11^T}\mQ - \frac{1}{N}\mQ{\bf 11^T} + \frac{1}{N^2}{\bf 11^T}({\bf 1}^T\mQ{\bf 1})
\end{equation}
\begin{equation}
    \text{recon + weight loss} \propto \Tr[(\lambda \mathds{1} + P(\mQ))^{-1}\bar\mX^T\bar\mX]
\end{equation}
Our previous convexity results then apply to the full similarity matrix.

\newpage

\section{Local Optimality Conditions}\label{sec_app:local_optimality_conditions}

We will now derive constraints that $\mZ$ must satisfy to be locally optimal. To do this we will perturb the representation and find conditions under which the first-order change to the loss is nonnegative. We will consider three types of perturbation, one completely general, and two that probe neuron-neuron and neuron-residual relationships.

\subsection{Preamble}

To begin, we want our perturbation to preserve nonnegativity, and we want to ignore an `uninteresting' kind of suboptimality. If any $\vz_d$ has a minimum value greater than zero the loss can be reduced by shifting this neuron's activity down. We assume this has been done so $\min \vz_d = 0$. To remove this possibility we parametrise the representation and the perturbation directly in terms of its mean subtracted version, then add the correct offset to preserve nonnegativity:
\begin{equation}
    \mZ \rightarrow \bar\mZ + \Delta - \min_{\text{row-wise}}[\bar\mZ + \Delta]{\bf 1^T} \qquad \qquad \bar\mZ \rightarrow \bar\mZ + \Delta
\end{equation}
Where $\Delta \in \mathbb{R}^{d_z \times N}$ is a row-mean-zero perturbation matrix.

The first-order perturbation to the weight-and-reconstruction term is then:
\begin{equation}
    -2 \Tr[(\lambda \mathds{1} + \bar\mZ^T\bar\mZ)^{-1}\bar\mZ^T\Delta(\lambda \mathds{1} + \bar\mZ^T\bar\mZ)^{-1}\bar\mX^T\bar\mX] 
\end{equation}
While the activity loss depends on the minimal de-meaned response of each neuron:
\begin{equation}
    {\bf 1}^T\mZ^T\mZ{\bf 1} = N^2\sum_d(\min[\bar\vz_d + \Delta_d])^2
\end{equation}
To first order, the change in minima for each neuron is:
\begin{equation}
    (\min[\bar\vz_d + \Delta_d])^2 - (\min[\bar\vz_d])^2 = (\min[\bar\vz_d] + \min_{i \in \mathcal{A}_d}[\Delta_{d}^{[i]}])^2 - (\min[\bar\vz_d])^2 = 2 \min[\bar\vz_d]\min_{i \in \mathcal{A}_d}[\Delta_{d}^{[i]}]
\end{equation}
where $\mathcal{A}_d$ is the set of datapoints at which $z^{[i]}_d = 0$, the inactive set. 

We introduce one further notational convenience, we denote with $m_d$ the L1 norm of the $d$th neuron:
\begin{equation}
    m_d = ||\vz_d||_1 = {\bf 1}^T\vz_d = N|\min[\bar\vz_d]|
\end{equation}
By the scaling arguments presented in~\cref{app_sec:symmetry_reformulations}, this is also equal to the L2 norm of the $d$th column of the dictionary: $||\vw_d||_2 = m_d$.

Hence, half the total first order change in the loss is:
\begin{equation}
    \frac{1}{2}\Delta\mathcal{L} = -\sum_d m_d N \min_{i \in \mathcal{A}_d}[\Delta_{d}^{[i]}]- \Tr[(\lambda \mathds{1} + \bar\mZ^T\bar\mZ)^{-1}\bar\mZ^T\Delta(\lambda \mathds{1} + \bar\mZ^T\bar\mZ)^{-1}\bar\mX^T\bar\mX] 
\end{equation}
Since the weight/reconstruction loss is differentiable and the activity penalty breaks down by neuron, we can consider a first-order perturbation to only one neuron (i.e. $\Delta$ has only one non-zero row). If the loss is unstable there must be a neuron-perturbation that would reduce the loss. Hence, we may instead consider the single neuron perturbation:
\begin{equation}
    \frac{1}{2}\Delta\mathcal{L}_d = -m_d N \min_{i \in \mathcal{A}_d}[\Delta_{d}^{[i]}]-\Delta_d^T(\lambda \mathds{1} + \bar\mZ^T\bar\mZ)^{-1}\bar\mX^T\bar\mX(\lambda \mathds{1} + \bar\mZ^T\bar\mZ)^{-1}\vz_d
\end{equation}
For writing convenience, using the definition of the optimal dictionary matrix, $\mW^*(\mZ) = \bar\mX(\lambda\mathds{1} + \bar\mZ^T\bar\mZ)^{-1}\bar\mZ^T$, we can identify the $d$th dictionary element, allowing us to clean this up slightly:
\begin{equation}\label{eq_app:change_in_loss_simple}
    \frac{1}{2}\Delta\mathcal{L}_d = -m_d N \min_{i \in \mathcal{A}_d}[\Delta_{d}^{[i]}]-\Delta_d^T(\lambda \mathds{1} + \bar\mZ^T\bar\mZ)^{-1}\bar\mX^T\vw_d
\end{equation}

\subsection{A General First-Order Perturbation}\label{sec_app:general_perturation}

%\will{Sharpen to show gap to necessity. Relate edge cases to existence of other solutions.}

If the change in loss,~\cref{eq_app:change_in_loss_simple}, is nonnegative for all mean-zero $\Delta_d$, $\mZ$ is first-order stable; if we can find one $\Delta_d$ that makes this negative it is first-order unstable. Let's see if such an instability inducing $\Delta_d$ exists.

Writing a shorthand for a scaling of the gradient of the reconstruction-and-weight loss:
\begin{equation}
    \vg_d = \frac{1}{Nm_d}(\lambda\mathds{1} + \bar\mZ^T\bar\mZ)^{-1}\bar\mX^T\vw_d 
\end{equation}
Our condition for instability is:
\begin{equation}
    \min_{i \in \mathcal{A}_d}[\Delta_{d}^{[i]}] > -\Delta_d^T\vg_d
\end{equation}
This condition can be equivalently expanded as:
\begin{equation}
    \Delta_{d}^{[i]} >  -\Delta_d^T\vg_d   \quad \forall i \in \mathcal{A}_d
\end{equation}
This is now a set of $|\mathcal{A}_d|$ linear conditions that we can stack as:
\begin{equation}
    (\mathds{1}_{\mathcal{A}_d} + {\bf 1}\vg_d^T)\Delta_d > 0 \qquad {\bf 1}^T\Delta_d = 0
\end{equation}
Where $\mathds{1}_{\mathcal{A}_d}\in \mathcal{R}^{|\mathcal{A}_d|\times N}$ is the matrix that extracts only the elements of $\Delta_d$ within $\mathcal{A}_d$, and the last constraint enforces $\Delta_d$ mean-zero.

This is a feasibility problem, if we can find a $\Delta_d$ that satisfies these constraints $\mZ$ is not a local minima. We can switch to the dual feasible problem \citep{boyd2004convex} and derive a dual certificate. This is another feasibility problem, only one of which has a solution. In particular, we will use Motzkin's Transposition Theorem \citep{motzkin1936beitrage}, which tells us a dual certificate is that there exists a solution to:
\begin{equation}\label{eq_app:dual_certificate}
    (\mathds{1}_{\mathcal{A}_d}^T + \vg_d{\bf 1}^T)\vy + {\bf 1}\mu = 0 \qquad \vy \geq  0, \qquad \vy\in\mathbb{R}^{|\mathcal{A}_d|}, \mu\in\mathbb{R}
\end{equation}
We can project these expressions onto the ${\bf 1}$ vector and, using the fact that $\vg_d$ is mean-zero, find:
\begin{equation}
    \mu = -\frac{1}{N}\sum_{j} y_j = -\frac{||\vy||_1}{N}
\end{equation}
Using this we get the following constraints on $\vg_d$:
\begin{equation}\label{eq_app:g_conditions}
    g^{[i]}_{d} = \begin{cases}
        \frac{1}{N} \qquad \qquad i \notin \mathcal{A}_d \\
        \frac{1}{N}
        - \frac{y_i}{||\vy||_1} \quad i\in\mathcal{A}_d
    \end{cases}
\end{equation}
This is an interesting set of conditions, telling us that, for $\mZ$ to be stable, the vector $\vg_d$ must be constant on the active set of neuron $d$ and smaller on the inactive set, $\mathcal{A}_d$. We can simplify by assuming, wlog, that $||\vy||_1 = 1$ (since this just corresponds to a positive rescaling of $\vy$ and $\mu$, something permitted while maintaining the same dual certificate,~\cref{eq_app:dual_certificate}). We will show that our more specific constraints can be understood as projections of this condition. 

We can rewrite it in one more clean fashion, using the definition of $\vg_d$:
\begin{equation}
    m_dN\vg_d = (\lambda\mathds{1} + \bar\mZ^T\bar\mZ)^{-1}\bar\mX^T\vw_d  = m_d{\bf 1} - m_dN\vy
\end{equation}
where, in an abuse of notation, we've used $\vy$ to denote both the original $|\mathcal{A}_d|$-dimensional vector, and its $N$ dimensional analogue that is padded with zeros for all indices not in $\mathcal{A}_d$. Now using the Woodbury Matrix Inversion formula: 
\begin{equation}
    (\lambda\mathds{1} + \bar\mZ^T\bar\mZ)^{-1} = \frac{1}{\lambda}\mathds{1} - \frac{1}{\lambda}\bar\mZ^T(\lambda\mathds{1} + \bar\mZ\bar\mZ^T)^{-1}\bar\mZ
\end{equation}
Plugging this into the expression and identifying the factors of $\mW = \bar\mX\bar\mZ^T(\lambda\mathds{1}+\bar\mZ\bar\mZ^T)^{-1}$:
\begin{equation}\label{eq_app:cleaner_g_cond}
    \bar\mX^T\hat\vw_d - \bar\mZ^T\mW^T\hat\vw_d = \lambda{\bf 1} - N\lambda \vy
\end{equation}
Where we've used optimal scaling to cancel $m_d$, replacing $\vw_d$ with its unit vector version $\hat\vw_d$. This will be a useful form in future.

%\paragraph{Uniqueness of Optima} Satisfying this condition is necessary for being a local optima, when is it sufficient? If the dual certificate $\vy > 0$, i.e. is strictly positive, the minima is sharp to first order, and so this becomes a sufficient condition.

\subsubsection{Deriving Locally Competitive Architecture}\label{app_sec:LCA}

Here we derive a standard ISTA-style result from this general condition. Specifically, we can develop an equation for $\bar\vz_d$. Starting from:
\begin{equation}
    \frac{1}{Nm_d}\bar\mX^T\vw_d = \frac{\lambda}{N}{\bf 1} - \lambda \vy - \bar\mZ^T\bar\mZ\vy
\end{equation}
 We can pull out the $\bar\vz_d$, which always equals $\min \vz_d$ when $\vy$ is nonzero:
\begin{equation}
    \bar\vz_d = \frac{1}{m_d}\bigg(\frac{1}{m_d}\bar\mX^T\vw_d - \lambda{\bf 1} + \lambda N \vy + N\sum_{d'\neq d}(\bar\vz_{d'}^T\vy)\bar\vz_{d'}\bigg)
\end{equation}
We can use the definition of $\vg_d$ again to clarify the $\vy^T\bar\vz_d$ term:
\begin{equation}
    \bar\vz_{d'}^T\vy = \bar\vz_{d'}^T(\vy - \frac{1}{N}{\bf 1}) =  -\bar\vz_{d'}^T\vg_d = \frac{-1}{Nm_d}\bar\vz_{d'}^T(\lambda\mathds{1} + \bar\mZ^T\bar\mZ)^{-1}\bar\mX^T\vw_d  = \frac{-1}{Nm_d}\vw_{d'}^T\vw_d
\end{equation}
Which tells us that:
\begin{equation}
    \bar\vz_d = \frac{1}{m_d}\bigg(\frac{1}{m_d}\bar\mX^T\vw_d - \lambda{\bf 1} + \lambda N\vy - \frac{1}{m_d}\sum_{d'\neq d}\vw_{d'}^T\vw_d\bar\vz_{d'}\bigg)
\end{equation}
Using the fact that $\vy$ is nonzero only when $\vz_d =0$, we can see this implements the ReLU operation, adding sufficient $\vy$ to take $\vz_d$ to 0. Combining this with adding the mean, $\frac{m_d}{N}$:
\begin{equation}
    \vz_d = \frac{1}{m^2_d}\text{ReLU}\bigg[\bar\mX^T\vw_d + m_d(\frac{m_d^2}{N}-\lambda){\bf 1}  - \sum_{d'\neq d}\vw_{d'}^T\vw_d\bar\vz_{d'}\bigg]
\end{equation}
We now use the optimal scaling property, \cref{app_sec:symmetry_reformulations} and denote with $\hat\vw_d$ the unit length version of the $d$th column of $\mW$ to find:
\begin{equation}
    \vz_d = \text{ReLU}\bigg[\frac{1}{m_d}\bar\mX^T\hat\vw_d + {\bf 1}\bigg(|\min \bar\vz_d| - \frac{\lambda}{m_d}\bigg) - \sum_{d'\neq d}\frac{m_{d'}}{m_d}\hat\vw_{d'}^T\hat\vw_d\bar\vz_{d'}\bigg]
\end{equation}
One can write this as a full population condition as follows:
\begin{equation}
    \mZ = \text{ReLU}\bigg[\underbrace{\text{diag}(\vm)^{-1}\hat\mW^T\bar\mX}_{\text{feedforward}} - \underbrace{\text{diag}(\vm)^{-1}\hat\mW^T\hat\mW\text{diag}(\vm)\bar\mZ_{/d}}_{\text{recurrent}} + \underbrace{\theta{\bf 1}^T}_\text{bias}\bigg]
\end{equation}
where we've introduced the bias vector $\theta_d = |\min \bar\vz_d| - \frac{\lambda}{m_d}$ and $\mZ_{/d}$ is $\mZ$ with the $d$th row set to 0.

This is essentially a rederivation of an ISTA fixed-point equation, as expected from theories of proximal optimality for L1-penalised problems \citep{hale2008fixed}. Precisely, when $||\vw_d||_2 = 1$ this corresponds exactly to the Locally Competitive Algorithm in \citep{rozell2008sparse}, and other works (e.g. \citep{pehlevan2014hebbian}). We can read from it three well-known architectural features of an optimal sparse decoder: an Affine+ReLU structure, lateral connections performing competitive inhibition, and weight tying.

There are two minor novel aspects of this derivation:
\begin{itemize}
    \item $\mW$ is defined entirely via the representation,~\cref{eq_app:optimal_dictionary_eqn}. Though we don't pursue it here, this naturally suggests using this self-consistent update as a dictionary learning algorithm.
    \item Our formulation recovers the weight tying in the encoder weights ($\hat\mW^T$ in the feedforward term), and the recurrent weight structure, $\hat\mW^T\hat\mW$, but demonstrates how to scale the encoder weights when they have non-unit-norms.
\end{itemize}

Further, we can read from this the familiar conditions under which a vanilla SAE,~\cref{eq_app:encoder_SAE}, recovers the optimal sparse setting: when the optimal weights are orthogonal. This result is somewhat dissatisfying, restricting the number of dictionary elements to $d_x$. There are slightly more relaxed settings in which decoupling the encoder and decoder weights, as is done in a vanilla SAE, is sufficient to recover the optimal sparse codes. For example, if the dot products between all features are nonpositive (as discussed in \citep{cui2026limits}), however this only gets you to maximally $2d_x$ features, and in general there remains a gap \citep{o2024compute}.

\subsection{Perturbation within Span of SAE Features}\label{sec_app:feature_feature_perturbation}

Now let's begin by assuming that the perturbation lives within the span of the SAE features:
\begin{equation}
    \Delta_d^T = \omega_d^T \bar\mZ \qquad \omega_d \in \mathbb{R}^{d_z}
\end{equation}
We derive our result in two equivalent ways. First, we restart from the perturbation equation, \cref{eq_app:change_in_loss_simple}, with this assumption. Second, we project the general conditions onto the span of the data, \cref{eq_app:g_conditions}.

\subsubsection{Via Subgradients}

The structure of the perturbation simplifies the change in loss, \cref{eq_app:change_in_loss_simple}:
\begin{align}
    \frac{1}{2}\Delta \mathcal{L} &= -m_d N\min_{i \in \mathcal{A}_d}[\sum_{d'}\omega_{d,d'}\bar z_{d'}^{[i]}] - \omega_d^T\bar\mZ(\lambda\mathds{1} + \bar\mZ^T\bar\mZ)^{-1}\bar\mX^T\vw_d  \\
     &= -m_d N \min_{i \in \mathcal{A}_d}[\sum_{d'}\omega_{d,d'}\bar z_{d'}^{[i]}] - \omega_d^T\mW^T\vw_d = \chi(\omega_d)
\end{align}
where, again, $\mathcal{A}_d$ is the set of datapoints for which $z^{[i]}_d = 0$. 

Each term inside the maximum is linear in $\omega_d$, so $\chi$ is a pointwise maximum of linear functions plus a linear term. It is therefore convex and positively homogeneous, with $\chi(\boldsymbol{0}) = 0$. As a result, the representation is stable to all such perturbations, $\chi(\omega_d)\ge 0$ for every $\omega_d$, if and only if $\omega_d=\boldsymbol{0}$ is a global minimiser of $\chi$, i.e.
\begin{equation}
\boldsymbol{0} \in \partial \chi(\boldsymbol{0}).
\end{equation}

To compute this subdifferential we use Danskin's theorem: the subdifferential of a
pointwise maximum of affine functions is the convex hull of the gradients of the
functions that attain the maximum,
\begin{equation}
f(\boldsymbol{x}) = \max_i f_i(\boldsymbol{x})
\quad\Longrightarrow\quad
\partial f(\boldsymbol{x}) = \mathrm{conv}\big\{\nabla f_i(\boldsymbol{x}) \,\big|\, f_i(\boldsymbol{x}) = f(\boldsymbol{x})\big\}.
\end{equation}
At $\omega_d = \boldsymbol{0}$ every $i\in\mathcal{A}_d$ attains the maximum, so
\begin{equation}
\partial \chi(\boldsymbol{0})
= -m_d N\,\mathrm{conv}\big\{\bar{\boldsymbol{z}}^{[i]} \,\big|\, i\in\mathcal{A}_d\big\}
  - \boldsymbol{W}^\top\boldsymbol{w}_d .
\end{equation}
The stability condition $\boldsymbol{0}\in\partial \chi(\boldsymbol{0})$ is therefore the inclusion
\begin{equation}
\frac{\boldsymbol{W}^\top\boldsymbol{w}_d}{-N m_d}
\in \mathrm{conv}\big\{\bar{\boldsymbol{z}}^{[i]} \,\big|\, i\in\mathcal{A}_d\big\}.
\end{equation}
If it holds, no perturbation within the span of the SAE features decreases the loss to
first order. If it fails, there is a direction $\omega_d$ with
$\Delta\mathcal{L}<0$, and the representation is not locally optimal. (If
$\boldsymbol{W}^\top\boldsymbol{w}_d/(-Nm_d)$ lies on the boundary of the hull, some
directions leave the loss unchanged to first order, and second-order terms decide.)
A bit of massaging gets us to the stated condition, \cref{eq:feature_feature_constraint}:
\begin{equation}
-\cos\boldsymbol{\theta}_d \in
\mathrm{conv}\left(\left\{
\frac{\bar{\boldsymbol{z}}^{[i]}}{|\min\bar{\boldsymbol{z}}_{d'}|}
\,\middle|\, i\in\mathcal{A}_d\right\}\right),
\end{equation}
where $\cos(\theta_{d,d'}) = \hat{\boldsymbol{w}}_d^\top\hat{\boldsymbol{w}}_{d'}$.

\subsubsection{Via Projection of General Condition}

Alternatively, we left-multiply \cref{eq_app:g_conditions} by $\bar\mZ$:
\begin{equation}
    \bar\mZ\vg_d = \frac{1}{Nm_d}\mW^T\vw_d = \bar\mZ(\frac{1}{N}\textbf{1} - \vy) = -\bar\mZ\vy
\end{equation}
Since the only non-zero elements of $\vy$ are those in $\mathcal{A}_d$, and $||\vy||_1 = 1$, the right hand side is exactly a vector lying in the convex hull and we recover \cref{eq:feature_feature_constraint}.

\subsection{Residual Constraints}\label{sec_app:feature_residual_perturbation}

We can spot the residuals hiding in \cref{eq_app:cleaner_g_cond}:
\begin{equation}\label{eq_app:residual_1}
    \bar\mX^T\hat\vw_d - \bar\mZ^T\mW^T\hat\vw_d = \mE^T\hat\vw_d = \lambda{\bf 1} - N\lambda \vy
\end{equation}
Where we've defined the residual matrix $\mE = \bar\mX - \mW\bar\mZ$. Let's take the dot-product of this with some mean-zero $N$-dimensional vector, $\delta \vz$:
\begin{equation}
     \hat\vw_d^T\mE\delta \vz = - N\lambda \delta \vz^T \vy
\end{equation}
And, recalling the properties of $\vy$, we recover \cref{eq:residual_constraint}. Easy!

\subsection{Second Order Condition - Resurrecting Neurons}\label{app_sec:resurrecting_neurons}

The general first-order condition, \cref{eq_app:g_conditions} is necessary for a solution, $\mZ$, to be a local minima, however it is not sufficient. There could be second order perturbations that decrease the loss. The clearest example of this is dead neurons waking up. They contribute a second order term to both the activity, weight, and reconstruction losses which is missed by our initial analysis. Here we derive a second order optimality condition from dead neurons waking up. Specifically, consider a representation whose $d$th neuron is inactive, then:
\begin{equation}
    \bar\mZ \rightarrow \bar\mZ + \ve_d \bar\vz_d^T
\end{equation}
We can calculate the change in the loss, \cref{eq:only_Z_dictionary_learning}, to second order in $\bar\vz_d$:
\begin{equation}
    N^2(\min \bar\vz_d)^2 - \bar\vz_d^T(\lambda \mathds{1} + \bar\mQ)^{-1}\bar\mX^T\bar\mX(\lambda \mathds{1} + \bar\mQ)^{-1}\bar\vz_d
\end{equation}
So we get a stability condition:
\begin{equation}
    \max_{\bar\vz_d:{\bf 1}^T\bar\vz_d=0}||\bar\mX(\lambda \mathds{1} + \bar\mQ)^{-1}\bar\vz_d||_2 \leq N |\min \bar\vz_d|
\end{equation}
This object is invariant to scaling up or down $\bar\vz_d$, so we can rescale to just consider cases where $\min \bar\vz_d \geq -1$. This makes it a maximisation over a polyhedron, the feasible set: $\{\bar\vz_d:{\bf 1}^T\bar\vz_d=0, \min\vz_d \geq -1\}$. The feasible set is a bounded polytope, since no element can be larger than $N-1$. The maxima of optimising a convex function over a polytope is achieved at one of the vertices, so we can just check for the vertex with highest value. The vertices are given by the settings in which all elements equal $-1$, except one, which equals $N-1$. Hence our condition becomes:
\begin{equation}
    \frac{1}{N}\max_{i}||\bar\mX(\lambda \mathds{1} + \bar\mQ)^{-1}(N\ve_i- {\bf 1})||_2 = \max_{i}||\bar\mX(\lambda \mathds{1} + \bar\mQ)^{-1}\ve_i||_2 \leq 1
\end{equation}
Which is nice and simple. An even cleaner result pops out if we use the Woodbury Matrix Inversion Lemma to finagle $\bar\mQ$:
\begin{equation}
    \bar\mX(\lambda \mathds{1} + \bar\mZ^T\bar\mZ)^{-1} = \frac{1}{\lambda}\bar\mX - \frac{1}{\lambda}\bar\mX\bar\mZ^T(\lambda\mathds{1} + \bar\mZ\bar\mZ^T)^{-1}\bar\mZ = \frac{1}{\lambda}(\bar\mX - \mW\bar\mZ)
\end{equation}
Which tells us that it all comes down to residuals!
\begin{equation}\label{eq_app:residual_wake_up_condition}
    \max_i ||\bar\vx^{[i]} - \mW\bar\vz^{[i]}||_2 \leq \lambda
\end{equation}

\newpage

\section{Specific Data Distributions}\label{app_sec:specific_data_distributions}

Here we ask when an encoding in which each neuron responds to a single source is stable, i.e.\ satisfies \cref{eq:feature_feature_constraint}. We first give a general sufficient condition for separable data. We then derive an exact condition for mutually exclusive categories, and ask when every choice of dictionary directions is stable.

We assume each neuron is sometimes off, $\min_i z^{[i]}_j = 0$ (else shift it down to improve the code). The demeaned activity then has $\min_i \bar z^{[i]}_j = -\mathbb{E}_i[z_j]$, and the normalised activity in \cref{eq:feature_feature_constraint} is:
\begin{equation}\label{eq:normalised}
\frac{\bar z^{[i]}_j}{|\min \bar z_j|} = \frac{z^{[i]}_j}{\mathbb{E}_i[z_j]} - 1 .
\end{equation}
This equals $-1$ wherever neuron $j$ is off and averages to zero over the data. Since the normalised activity is at most $\max_i z_j/\mathbb{E}_i[z_j] - 1$, a necessary condition for \cref{eq:feature_feature_constraint} is
\begin{equation}\label{eq:necessary}
1-\cos\theta_{k,j} \le \frac{\max_i z^{[i]}_j}{\mathbb{E}_i[z_j]} \qquad \forall\, j\neq k .
\end{equation}

\paragraph{Dictionary versus embedding geometry.} The cosines in \cref{eq:feature_feature_constraint}, $\cos\theta_{k,j} = \hat{\boldsymbol{w}}_k^T\hat{\boldsymbol{w}}_j$, are between the learnt dictionary elements, not between the directions along which sources are embedded in the data. The two are closely related. Suppose the data are generated as $\boldsymbol{x}^{[i]} = \sum_k \boldsymbol{u}_k s^{[i]}_k$, and stack the $\boldsymbol{u}_k$ as the columns of $\boldsymbol{U}$ and the sources as the rows of $\boldsymbol{S}$. A single-source code, $z^{[i]}_k = a_k s^{[i]}_k$ with $a_k > 0$, has optimal dictionary:
\begin{equation}\label{eq:dictionary_embedding}
\boldsymbol{W}^* = \bar{\boldsymbol{X}}(\lambda\mathds{1}+\bar{\boldsymbol{Z}}^T\bar{\boldsymbol{Z}})^{-1}\bar{\boldsymbol{Z}}^T
= \boldsymbol{U}\,(\lambda\mathds{1} + \boldsymbol{\Sigma}\boldsymbol{A}^2)^{-1}\boldsymbol{\Sigma}\boldsymbol{A},
\end{equation}
where $\boldsymbol{A} = \mathrm{diag}(a_k)$ and $\boldsymbol{\Sigma} = \bar{\boldsymbol{S}}\bar{\boldsymbol{S}}^T$ is the (unnormalised) source covariance. If $\boldsymbol{\Sigma}$ is invertible, then as $\lambda\to 0$, $\boldsymbol{W}^*\to\boldsymbol{U}\boldsymbol{A}^{-1}$, so each $\hat{\boldsymbol{w}}_k\to\hat{\boldsymbol{u}}_k$: for small $\lambda$ the dictionary geometry is the embedding geometry. At finite $\lambda$, correlated sources mix the dictionary elements slightly. For example, mutually exclusive sources have negative covariance, so orthogonally embedded categories give a nearly, but not exactly, orthogonal dictionary. We therefore state all results in terms of the dictionary cosines. Whenever a condition holds strictly for an orthogonal dictionary, it also holds, by continuity, for orthogonally embedded sources at small enough $\lambda$.

\subsection{Separable data}

The classic sparse identifiability setting models data as a sparse combination of
sources, $\boldsymbol{x}^{[i]} = \sum_{k=1}^K \boldsymbol{u}_k s^{[i]}_k$, and asks
when a dictionary in which each element encodes a single source,
$z^{[i]}_k \propto s^{[i]}_k$, is optimal. One classic, if limiting, assumption from
the matrix factorisation literature is separability \citep{donoho2003does}: for
each source $j$ there is a datapoint $i_j$ at which only source $j$ is active.

Suppose the data are separable, and there is also a datapoint at which no source is
active. We will show that the single-source encoding satisfies \cref{eq:feature_feature_constraint}
whenever, for every $k$,
\begin{equation}\label{eq:sep_condition}
\sum_{j\neq k} \big(1-\cos\theta_{k,j}\big)\,\frac{\mathbb{E}_i[z_j]}{z^{[i_j]}_j} \;\le\; 1 .
\end{equation}

To see this, consider a particular feature $k$. The datapoints in which each other feature is active alone, and the datapoint in which all features are off, all have $z_k = 0$, so they appear in the convex hull for neuron $k$. By
\eqref{eq:normalised}, the all-off datapoint normalises to $-\boldsymbol{1}$, and $i_j$ normalises
to $-\boldsymbol{1} + \big(z^{[i_j]}_j/\mathbb{E}_i[z_j]\big)\boldsymbol{e}_j$. Take the
weights
\begin{equation}
\lambda_j = \big(1-\cos\theta_{k,j}\big)\frac{\mathbb{E}_i[z_j]}{z^{[i_j]}_j}\quad (j\neq k),
\qquad
\lambda_0 = 1-\sum_{j\neq k}\lambda_j .
\end{equation}
These are nonnegative, since $\cos\theta_{k,j}\le 1$ and assuming \eqref{eq:sep_condition},
and they sum to one. The corresponding convex combination is
\begin{equation}
-\boldsymbol{1} + \sum_{j\neq k}\lambda_j\frac{z^{[i_j]}_j}{\mathbb{E}_i[z_j]}\,\boldsymbol{e}_j
= -\boldsymbol{1} + \sum_{j\neq k}\big(1-\cos\theta_{k,j}\big)\boldsymbol{e}_j
= -\cos\boldsymbol{\theta}_k .
\end{equation}

For an orthogonal dictionary, $\cos\theta_{k,j}=\delta_{kj}$, and \eqref{eq:sep_condition}
reads $\sum_{j\neq k}\mathbb{E}_i[z_j]/z^{[i_j]}_j \le 1$. Each source must be small
on average compared with its value when it is the only active source. If this holds strictly, orthogonally embedded sources are stable at small $\lambda$. Further
datapoints, such as those in which several sources are active, only enlarge the
convex hull. So \eqref{eq:sep_condition} is sufficient but not in general necessary.

\subsection{Mutually exclusive categories}

Consider a categorical variable that takes one of $K$ values, value $d$ with
probability $q_d$, or is inactive with probability $q_0 = 1-\sum_d q_d$. We encode it
as a $K$-dimensional one-hot or zero vector $\boldsymbol{x}^{[i]}$, and give each category its own neuron, $z^{[i]}_d \propto x^{[i]}_d$. Then $\mathbb{E}_i[z_d]/z_d = q_d$ whenever category $d$ is active. We will show that if $q_0 > 0$, the one-hot encoding satisfies \cref{eq:feature_feature_constraint} if and
only if, for every $k$,
\begin{equation}\label{eq:mutex}
\sum_{j\neq k} q_j\big(1-\cos\theta_{k,j}\big) \;\le\; 1 ,
\end{equation}
while if $q_0 = 0$, it is unstable whenever the dictionary elements are orthogonal or positively aligned.

To see this, again consider a feature $k$. The datapoints with $z_k=0$ are the inactive ones, which normalise to
$-\boldsymbol{1}$, and those in which some other category $j\neq k$ is active, which normalise to
$-\boldsymbol{1} + \boldsymbol{e}_j/q_j$. A convex combination with weight $\lambda_j$
on category $j$ and $\lambda_0$ on the inactive points is
$-\boldsymbol{1} + \sum_{j\neq k}(\lambda_j/q_j)\boldsymbol{e}_j$. This equals
$-\cos\boldsymbol{\theta}_k$ if and only if $\lambda_j = q_j(1-\cos\theta_{k,j})$. These
weights are nonnegative, and a nonnegative $\lambda_0$ completes them to sum to one
exactly when \eqref{eq:mutex} holds.

If $q_0 = 0$ there are no inactive datapoints, so the weights must satisfy $\sum_{j\neq k}\lambda_j = 1$ exactly. When every $\cos\theta_{k,j}\ge 0$, this sum is at most $\sum_{j\ne k} q_j = 1-q_k < 1$, so the condition fails. (In fact, with $q_0=0$ the demeaned one-hot vectors sum to zero, so $\boldsymbol{\Sigma}$ is singular and, by \cref{eq:dictionary_embedding}, the dictionary elements are linearly dependent and can never be orthogonal.) For an orthogonal dictionary with $q_0 > 0$, the left-hand side of \eqref{eq:mutex} is $1-q_0-q_k < 1$, so the one-hot code is stable. Since this inequality is strict, the same holds for orthogonally embedded categories, such as the one-hot data itself, at small enough $\lambda$.

\begin{comment}

\subsection{Independent features}

At the other extreme, suppose each source is binary, $z_j\in\{0,1\}$ up to scale, and
active with probability $p_j$, and that every combination of active sources occurs in
the data.

\begin{proposition}\label{prop:independent}
If every combination of active sources occurs, the single-source encoding satisfies
\eqref{eq:stability_condition} if and only if
\begin{equation}\label{eq:indep}
p_j\big(1-\cos\theta_{k,j}\big) \le 1 \qquad \forall\, j\neq k .
\end{equation}
\end{proposition}

\begin{proof}
Necessity is \eqref{eq:necessary}, since $\max_i z_j/\mathbb{E}_i[z_j] = 1/p_j$. For
sufficiency, fix $k$ and set $\pi_j = p_j(1-\cos\theta_{k,j})\in[0,1]$. Weight each
datapoint with $z_k=0$ and active set $S$ by the probability that $S$ is drawn when
each $j\neq k$ is included independently with probability $\pi_j$. These weights are
convex, and component $j$ of the combination is
$-1 + \pi_j/p_j = -\cos\theta_{k,j}$.
\end{proof}

\end{comment}

\subsection{When is every geometry stable?}

The previous results found simple conditions that ensured features with orthogonal dictionary elements were stable. Here we ask when every possible dictionary geometry is stable. Each condition above constrains the cosines of one neuron $k$ at a time, and becomes harder to satisfy as any $\cos\theta_{k,j}$ decreases. The worst case is $\cos\theta_{k,j} = -1$ for all $j\neq k$, i.e.\ $\hat{\boldsymbol{w}}_j = -\hat{\boldsymbol{w}}_k$. This geometry arises from real data: embed every other source anti-aligned with the $k$-th, $\boldsymbol{u}_j = -\boldsymbol{u}_k$. Then $\boldsymbol{U}$ has rank one, so by \cref{eq:dictionary_embedding} every dictionary element is parallel or anti-parallel to $\boldsymbol{u}_k$, and for small enough $\lambda$ (with $\boldsymbol{\Sigma}$ invertible) the signs match those of the embedding. So the encoding is stable for every dictionary geometry if and only if the condition holds with every other dictionary element anti-aligned with $k$. In the two cases, these conditions are:

\begin{itemize}
\item \textbf{Mutually exclusive} (exact):
\begin{equation}\label{eq:mutex_all}
\sum_{j\neq k} q_j \le \tfrac12 \;\;\forall k
\quad\Longleftrightarrow\quad
q_0 + \min_k q_k \;\ge\; \tfrac12 .
\end{equation}
For $K$ equally likely categories, $q_d = (1-q_0)/K$, this is
$q_0 \ge \frac{K-2}{2(K-1)}$, which tends to $\tfrac12$ as $K$ grows. The more often
the variable is off, the wider the range of stable geometries, until every geometry is
stable once it is off about half the time.
\item \textbf{Separable} (sufficient): $\sum_{j\neq k}\mathbb{E}_i[z_j]/z^{[i_j]}_j \le \tfrac12$ for every $k$.

Writing $\mathbb{E}_i[z_j] = p_j\,\mathbb{E}_i[z_j \mid z_j>0]$, each term is the frequency $p_j$ of source $j$, scaled by its typical active value relative to its value at the separating datapoint. For sources whose separating value, $z_j^{[i_j]}$ is typical the condition becomes $\sum_{j\neq k}p_j\le\tfrac12$, matching the mutually exclusive case. 
\end{itemize}

In both cases, positively aligned dictionary elements are easy to accommodate, and the binding
constraint comes from elements anti-aligned with one another. For mutually exclusive categories, the other categories must \emph{jointly} be rare.

\subsection{Toy Test Experiments}\label{sec_app:feature_experiments}

We train a two-element dictionary on the three datasets shown in \cref{fig:feature_fig}A-C, and measure how mixed the resulting representation is using an amodularity index,
\begin{equation}
\text{Amodularity} = \frac{1}{2}\sum_{d=1}^{2}\min_{k\in{1,2}} \big|\rho(z_d, s_k \mid s_{\neg k})\big|,
\end{equation}
where $\rho(z_d, s_k \mid s_{\neg k})$ is the partial correlation between neuron $d$'s activity and source $k$ after regressing out the other source. The index is zero when each neuron's activity depends on only one source, and increases as neurons mix the two.

\section{Real SAE experiments}\label{app_sec:real_SAE_experiments}

\subsection{Setup}
We use the SAE-lens releases \texttt{gpt2-small-res-jb} (24,576 latents per layer) and \texttt{gpt2-small-res-jb-feature-splitting} (768, 1,536, 3,072 and 6,144 latents, layer 8 only). All were trained on the \texttt{hook\_resid\_pre} activations of GPT2-small, on OpenWebText, with the same recipe and L1 coefficient ($8\times10^{-5}$), and with unit-norm decoder rows. Unless stated otherwise, we use the 24,576-latent layer-8 SAE.

For each dataset we take 512 contexts of 128 tokens, which gives 65,024 tokens after dropping BOS positions (whose huge-norm residuals are outliers). Documents are shuffled and truncated to 4,000 characters, so each sample spans many documents. The datasets are:
\begin{itemize}
\item OpenWebText (two disjoint samples);
\item C4; English Wikipedia; the Pile;
\item German and French Wikipedia (one parquet shard each);
\item code (\texttt{NeelNanda/code-10k}), and Python code (the same, restricted to \texttt{.py} files).
\end{itemize}

\subsection{The pairwise test and the hull gauge}
Write $u_d = z_d/\mathrm{mean}(z_d)$. This equals $\bar z_d/|\min \bar z_d| + 1$ when $\min z_d = 0$, and is exactly 0 when the feature is off. ~\cref{eq:feature_feature_constraint} projected onto a set $S$ of other features reads
\[ 1-\cos(\hat w_d,\hat w_S) \in \mathrm{conv}\{u_S^{[i]} : z_d^{[i]}=0\}. \]
Tokens at which $d$ and every feature in $S$ are off sit at the origin. Provided such a token exists (which is always the case in our data), the condition is therefore equivalent to the \emph{hull gauge}
\[ g(d,S) = \min\Big\{\textstyle\sum_i y_i \,:\, y\ge 0,\ \sum_{i:z_d^{[i]}=0} y_i\, u_S^{[i]} = 1-\cos(\hat w_d,\hat w_S)\Big\} \le 1, \]
which is a linear program. $g=\infty$ means the target lies outside the cone of the points, e.g.\ a feature in $S$ never fires while $d$ is off: a strict hierarchy. For a single feature, $g=(1-\cos\theta_{dd'})/\max_{A_d} u_{d'}$, which is the pairwise test. Because it is a maximum, a single token with the child active and the parent off can be enough to pass. A pair can also fail at the lower end of the interval, when the two features are never off together (fig.~1B); this never occurs in our data.

We test all ordered pairs of features active on more than 0.5\% of tokens (at least $\sim$325 activations). Lowering the cutoff to 0.2\% (8,800 features, 77M pairs) also gives no violations on OWT.

\subsection{Layer shift}
We run GPT2-small on the same OWT tokens and apply the layer-8 SAE, unchanged, to the residual stream at layers 4, 6, 7, 9 and 10. Its fraction of variance unexplained is 0.77, 0.47, 0.30, 0.25 and 1.38 respectively, against 0.14 at layer 8. The number of violating pairs, all strict hierarchies, is 146, 22, 0, 0 and 3. Every violation replicates in both halves of the tokens. Because the tokens are identical, each violating pair can be compared directly with the same pair at layer 8:
\begin{itemize}
\item \textbf{Layer 6.} 18 of the 22 violations share one parent (16021). At layer 8 it fires on 6--38\% of tokens in the first 40 positions of a context, and almost never later; at layer 6 it fires on 100\% of them. Its children are features that also fire only early in the context, and at layer 8 they fire mostly while the parent is off (e.g.\ 951 of 1,369 activations).
\item \textbf{Layer 4.} The mirror image: parent 4760 fires on 9--22\% of late tokens at layer 8, and on 99--100\% of tokens beyond position 40 at layer 4, with late-position children nested inside it.
\item \textbf{Layer 10.} Parent 5402 goes from 1--3\% of tokens at layer 8 to 84--99\% at every position.
\end{itemize}
The tokens on which the parents are off are ordinary tokens (``the'', commas, ``of''), with as many other features active as elsewhere, so the effect is not an artefact of inactive tokens.

\subsection{Joint violations}
The joint search is not exhaustive. As targets $d$ (the feature whose inactive tokens form the hull), we take only the 25 features whose largest pairwise gauge is highest, i.e.\ those whose pair constraints are already closest to failing. For each, we test the joint condition against the $k$ features that individually come closest to violating ($k=1,3,10,30,100$), excluding features that already violate pairwise. For every violating set, we greedily remove features while the set still violates, which gives a minimal violating set. Results:
\begin{itemize}
\item OWT, C4, English Wikipedia and the Pile: no set violates; the median gauge at $k=100$ is 0.19--0.26.
\item German and French: no set violates, although gauges are the highest of any dataset (median at $k=100$: 0.60 and 0.57).
\item Code: 2 of the 25 targets violate, with minimal sets of 20--25 features; the worst pair within each set has gauge $\le 0.14$.
\item Python code: 3 of 25 violate, with minimal sets of 21--80 features.
\end{itemize}
The joint gauge exceeds the pairwise ones because the members' activity outside the parent comes from largely disjoint tokens, and the weights must sum to at most 1. When there is no overlap, the joint gauge is the sum of the pairwise gauges; for one code set the pairwise gauges sum to 1.8, while the joint gauge is 1.09. In the code sets, the target fires on essentially every indentation space, and the members are features that each fire at a different depth within the run of indentation spaces.

\subsection{What the violating features do on OWT}
For each dataset violation, we look up the same features on 130k OWT tokens (both samples):
\begin{itemize}
\item \textbf{German and French.} The parents respond to foreign words inside English text (``Le Monde Diplomatique'', ``r\'esum\'e''). The children are generic English grammar features (``that / which / who''; ``this / that''; ``as / like''; sentence-final newlines; commas), and each fires almost always while the parent is off (e.g.\ 2,115 of 2,129 activations, and 626 of 626).
\item \textbf{Python.} The parent fires on stray space tokens in prose. The child fires on islands and beaches (``Sapelo Island''), never with the parent on.
\end{itemize}
The hierarchies are therefore not present in the SAE's training data. They arise because, off-distribution, one feature saturates and others fire only within it.

\subsection{Feature-residual condition}
Let $E$ be the demeaned residuals, $u$ a principal direction of $E$ with variance $\sigma^2$, and $s=E^\top u$. Eq.~(7) with $\delta z = s$ gives, for any residual,
\[ -\sigma^2 \cos(\hat w_d,u) \in \lambda\,[\min_{A_d} s,\ \max_{A_d} s], \]
the full-rank version of eq.~(8). For each feature $d$ and each of the top 50 residual directions, we compute the smallest $\lambda$ at which the pair passes:
\[ \lambda_{\min}(d,u) = \sigma^2|\cos| \,/\, \big(\max_{A_d} s \text{ if } \cos<0, \text{ else } |\min_{A_d}s|\big), \]
and set $\lambda^* = \max \lambda_{\min}$. The condition holds for every $\lambda\ge\lambda^*$, so the SAE's residuals are consistent with its own training strength exactly when the residual break ratio $\lambda^*/\lambda_{\mathrm{train}}$ is at most 1.

\textbf{Training $\lambda$.} The SAEs were trained with the per-element MSE, normalised by the centred input norm, plus $c\sum_d|z_d|$ with $c=8\times10^{-5}$. In the convention of eq.~(1), this is $\lambda_i = c\, d_{\mathrm{in}}\,\|x_i-\bar x\|/2$, with median 2.32 on OWT. As an independent check, the median of $\hat w_d^\top e$ over each feature's active tokens, which equals $\lambda$ at an optimum (eq.~(48)), is 2.38 (5--95\% range across features: 2.07--2.70).

\textbf{Results.}
\begin{itemize}
\item OWT, C4, English, German and French Wikipedia: 0.19--0.35, at every width. The largest residual directions on German and French are larger than on OWT (standard deviation 4--9 vs.\ $\sim$3), but none is aligned with a feature in a violating way.
\item The Pile: 0.45--0.81.
\item Code: 1.0--1.9. The violation is borderline for the main SAE (4 features, exceeding the threshold by 1--5\%) and clear for the 768-latent SAE (18 features, by up to $1.9\times$). All violating features are aligned with residual PC1, which runs from indentation spaces to identifier fragments. On code, 60--100\% of their activations are on whitespace. On OWT most respond to formatting and byte-level tokens (GPT-2's byte fragments of curly quotes and apostrophes, e.g.\ in ``don't''; separators such as ``===='', and stray spaces).
\item Layer shift (main SAE): 0.78, 0.70, 0.63 and 0.52 at layers 4, 6, 7 and 9, so no violations. At layer 10 the value is 17, but the SAE no longer reconstructs the input there (FVU 1.38; residual standard deviation 66 vs.\ $\sim$3 at layer 8).
\end{itemize}

\section{Optimal Antipodal Dense Encodings}\label{sec_app:dense_encodings}

It is hard to solve the dictionary learning problem analytically. Here, we consider a toy case in which we {\it can} describe everything, which we use to understand how an SAE encodes dense variables,~\cref{sec:dense_antipodal_feature_pairs}. We consider an SAE trained on a scalar input. We find that with nonnegativity this leads to antipodal encoding of dense features - i.e. optimally the representation contains two neurons encoding the positive and negative component of the scalar. This turns out to be a special case of the wide-convex limiting solution considered in \cref{app_sec:more_neurons_than_datapoints}, except, thanks to the structure of the discovered solutions, the encoder architecture constraint is automatically satisfied, so in this case it is also a statement about vanilla SAEs.

\subsection{KKT Conditions}

In this analysis we will begin by deriving the KKT conditions for~\cref{eq_app:only_Z_dictionary_learning}. We enforce the nonnegativity of $\mZ$ with a matrix of Lagrange multipliers, $\mP$:
\begin{equation}
    \mathcal{M} = \textbf{1}^T\mZ^T\mZ\textbf{1} + \Tr[(\lambda\mathds{1} + \bar\mZ^T\bar\mZ)^{-1}\bar\mX^T\bar\mX] - 2\Tr[\mP^T\mZ]
\end{equation}
Taking the derivative with respect to $\mZ$ gives:
\begin{equation}
    \frac{1}{2}\frac{\partial \mathcal{M}}{\partial \mZ} = \mZ{\bf 1 1}^T - \bar\mZ(\lambda\mathds{1} + \bar\mZ^T\bar\mZ)^{-1}\bar\mX^T\bar\mX(\lambda\mathds{1} + \bar\mZ^T\bar\mZ)^{-1}(\mathds{1} - \frac{1}{N}{\bf 1 1}^T) - \mP
\end{equation}
Further, it is not hard to see that ${\bf 1}$ is an eigenvector of $(\lambda\mathds{1} + \bar\mZ^T\bar\mZ)^{-1}$, and, hence, since $\bar\mX$ is demeaned:
\begin{equation}
    \bar\mX(\lambda\mathds{1} + \bar\mZ^T\bar\mZ)^{-1}{\bf 1} \propto \bar\mX{\bf 1} = 0
\end{equation}
Which allows us to drop the term that ends ${\bf 11}^T$. Hence our final KKT conditions are:
\begin{equation}
    \mZ{\bf 1 1}^T - \bar\mZ(\lambda\mathds{1} + \bar\mZ^T\bar\mZ)^{-1}\bar\mX^T\bar\mX(\lambda\mathds{1} + \bar\mZ^T\bar\mZ)^{-1} - \mP = {\bf 0}, \qquad \mZ \geq 0, \quad \mP \geq 0, \quad \mZ\odot \mP = {\bf 0}
\end{equation}
Where $\odot$ denotes elementwise multiplication. In general these conditions are difficult to manipulate, but we shall make progress in the case of a scalar input.

\subsection{Smallest Optimal Scalar Encoding Uses (at most) 2 Neurons}

First, we shall show that there are, at most, two unique neural responses, $\vz_d$, in the population.

If each datapoint is a scalar, $x^{[i]}$, we can spot a single vector is repeated in the KKT conditions: $\va = (\lambda\mathds{1} + \bar\mZ^T\bar\mZ)^{-1}\bar\vx$. Using this vector, the KKT conditions become:
\begin{equation}
    \mZ{\bf 1 1^T} - \bar\mZ\va\va^T = \mP
\end{equation}
Breaking this down by neuron:
\begin{equation}\label{eq:neuron_wise_KKT}
    \vz_d^T{\bf 1 1} - \bar\vz_d^T\va \va = \mu_d{\bf 1} - \alpha_d \va = \vp_d
\end{equation}
This is great, because each neuron contributes only through two scalars, $\mu_d = \vz_d^T{\bf 1}$ and $\alpha_d = \bar\vz_d^T\va$. Otherwise everything is governed by the shared vector, $\va$. The sharing of this vector across the population places heavy constraints. Call the set of indices at which neuron $d$ has zero activity $\mathcal{A}_d$. Then, for all $i \notin \mathcal{A}_d$, $p_d^{[i]} = 0$ by the KKT conditions and:
\begin{equation}
    a^{[i]} =  \frac{\mu_d}{\alpha_d}
\end{equation}
Hence, the vector $\va$ is broken down into parts. Only in those parts whose elements equal the fixed value $\frac{\mu_d}{\alpha_d}$ can neuron $d$ have non-zero activity. Any other neuron that has a different value of $\frac{\mu_d}{\alpha_d}$ must then be off during these datapoints.

Now consider a second neuron, index $d'$, with its own set of inactive datapoints, $\mathcal{A}_{d'}$. Unless $\frac{\mu_{d'}}{\alpha_{d'}} = \frac{\mu_d}{\alpha_d}$, $\mathcal{A}_{d'}^c$ and $\mathcal{A}^c_d$, the complements of the two sets where the neurons are active, must be disjoint; let's consider the case where they are disjoint. Consider a datapoint $i \notin \mathcal{A}_d$, the other neuron must be off, $z_{d'}^{[i]} = 0$, therefore $p_{d'}^{[i]} \geq 0$. This tells us that:
\begin{equation}
    p_{d'}^{[i]} = \mu_{d'} - \alpha_{d'} a^{[i]} = \mu_{d'} - \alpha_{d'} \frac{\mu_{d}}{\alpha_{d}} \geq 0
\end{equation}
Considering the converse points, where neuron $d$ is off and $d'$ is on, we derive similarly that:
\begin{equation}
    \mu_d - \frac{\mu_{d'}\alpha_d}{\alpha_{d'}} \geq 0
\end{equation}
We know both $\mu_d$ and $\mu_{d'}$ are positive (since $\vz_d \geq 0$ and $\mu_d = {\bf 1}^T\vz_d$). Consider the case when $\alpha_d$ and $\alpha_{d'}$ have the same sign, then the only way for these two inequalities to be satisfied is if $\frac{\mu_d}{\alpha_d} = \frac{\mu_{d'}}{\alpha_{d'}}$, i.e. the neurons in fact share the same part of $\va$ vector. Alternatively they can have a different sign, and the two inequalities are satisfied. Extending this argument to more neurons we see they'll always fall into these two groups.

These groups don't have to be exactly one neuron. For example, the population could contain two neurons that were scaled copies of one another, or they could split the datapoints between them, with one neuron fitting one part, another neuron the rest. However, the optimal solution can be reached using any population with at least two neurons.

\subsection{Precise Representation}\label{sec_app:precise_dense_antipodal_representation}

Now we shall derive the functional form of the two-neuron optimal representation.

The definition of $\va$ tells us that: 
\begin{equation}\label{eq:data_link_eqn}
    \lambda \va + \alpha_+\bar\vz_+ + \alpha_-\bar\vz_- = \bar\vx
\end{equation}
This tells us, first, that ${\bf 1}^T\va = 0$. We can then pin down the remaining $\{a_\pm, \mu_\pm, \alpha_\pm\}$ a little more: 
\begin{equation}
    \alpha_+ = \va^T\bar\vz_+ = \va^T\vz_+ = a_+\sum_iz_+^{[i]}
\end{equation}
\begin{equation}
    \mu_+ = {\bf 1}^T\vz_+ = \sum_iz_+^{[i]}
\end{equation}
Hence:
\begin{equation}
    \alpha_+ = a_+ \mu_+ \qquad \alpha_+^2 = \mu_+^2
\end{equation}
Identical arguments work for $\mu_-$ and $\alpha_-$, and since we know that $\alpha_\pm$ have to have different signs, we end up with:
\begin{equation}
    a_+ = 1 = \frac{\mu_+}{\alpha_+} \quad a_- = -1 = \frac{\mu_-}{\alpha_-}
\end{equation}
Next, let's relate the representation back to the data. Using \cref{eq:data_link_eqn}:
\begin{align}
    z_+^{[i]} = \bigg[\frac{1}{\mu_+}(\bar x^{[i]} - \lambda - \frac{\mu_-^2 - \mu_+^2}{N})\bigg]_+ \\
    z_-^{[i]} = \bigg[
        \frac{1}{\mu_-}(-\bar x^{[i]} - \lambda + \frac{\mu_-^2 - \mu_+^2}{N}) \bigg]_+
\end{align}
This is a pretty implicit representation of $\mu_\pm$. We use the following self-consistency relation to numerically find the `effective threshold' $\theta$, using $\mu_\pm = {\bf 1}^T\vz_\pm$:
\begin{equation}\label{eq:implicit_theta}
    \theta = \frac{\mu_-^2 - \mu_+^2}{N} = \mathbb{E}_i\bigg[[-\bar x^{[i]} - \lambda + \theta]_+ - [\bar x^{[i]} - \lambda - \theta]_+  \bigg]
\end{equation}
So we create an array of $\theta$ values, and choose the value that minimises the difference between the left and right hand side. Having found a $\theta$ value, $\mu_\pm$ can be easily calculated from using:
\begin{equation}
    \mu_\pm^2 = {\bf 1}^T[ \pm \bar x^{[i]} - \lambda \mp \theta]_+
\end{equation}
Unless $\lambda$ is chosen very particularly, there will be no datapoint that lies exactly on the boundary, ensuring that strict complementary slackness applies, and hence making this solution the global optima. This completes our optimal modular representation.

\subsection{When is it one neuron?}

Our final question is where is the threshold between using one or two neurons? We shall answer this by proposing an optimal single neuron encoding, and finding when it is unstable to the addition of an extra neuron.

Consider the following optimal single neuron encoding, of the form identified in the previous section:
\begin{equation}
    \vz = a[\bar\vx - \theta{\bf 1}]_+ = a(\bar\vx - \theta{\bf 1} + \Lambda)
\end{equation}
Where $\Lambda$ is a vector that implements the ReLU, i.e.:
\begin{equation}
    \Lambda^{[i]} = \begin{cases}
        0 \quad \text{if} \quad \bar x^{[i]} - \theta >0\\
        -(\bar x^{[i]} - \theta)\quad \text{else}
    \end{cases}
\end{equation}
The mean-zero version is then:
\begin{equation}
    \bar\vz = a(\bar \vx + \Lambda - \frac{1}{N}{\bf 1}^T\Lambda {\bf 1})
\end{equation}
Now we can plug this into the first order optimality condition, \cref{eq_app:cleaner_g_cond}, to get:
\begin{equation}
    \bar\vx + N\lambda \vy - \lambda {\bf 1} = \bar\vz w =wa(\bar \vx + \Lambda - \frac{1}{N}{\bf 1}^T\Lambda {\bf 1})
\end{equation}
Hence, simply by equating factors, we find:
\begin{equation}
    wa = 1, \qquad \Lambda = N\lambda \vy
\end{equation}
Summing the indices on the second equation tells us:
\begin{equation}
    \frac{1}{N}\sum_i [-(\bar x^{[i]}- \theta)]_+ = \lambda
\end{equation}
Let's assume $\lambda$ is smaller than the distance between points, then this really constrains how far $\bar x^{[i]}$ can travel below the threshold!

Let's make a simple assumption that it is only below threshold when $\bar x^{[i]} = \min \bar \vx$. Then this is telling us that:
\begin{equation}
    \theta = \frac{\lambda}{q}- |\min \bar\vx|
\end{equation}
where $q$ is the proportion of points at which $x^{[i]} = \min \bar\vx$.

Finally, let's measure when this is stable to the addition of a new neuron. Using \cref{eq_app:residual_wake_up_condition} we see that a condition is that:
\begin{equation}
    |\min \bar \vx - wa(\theta -  \lambda)|= \lambda \frac{1-q}{q} \leq \lambda
\end{equation}
Hence $q > \frac{1}{2}$, the variable is at least half-sparse.

For larger $\lambda$ this condition will change slightly, in ways that could be calculated using a more elaborate version of this analysis.

\subsection{Empirical Verification}

\begin{figure}[h]
    \centering
    \includegraphics[width=\linewidth]{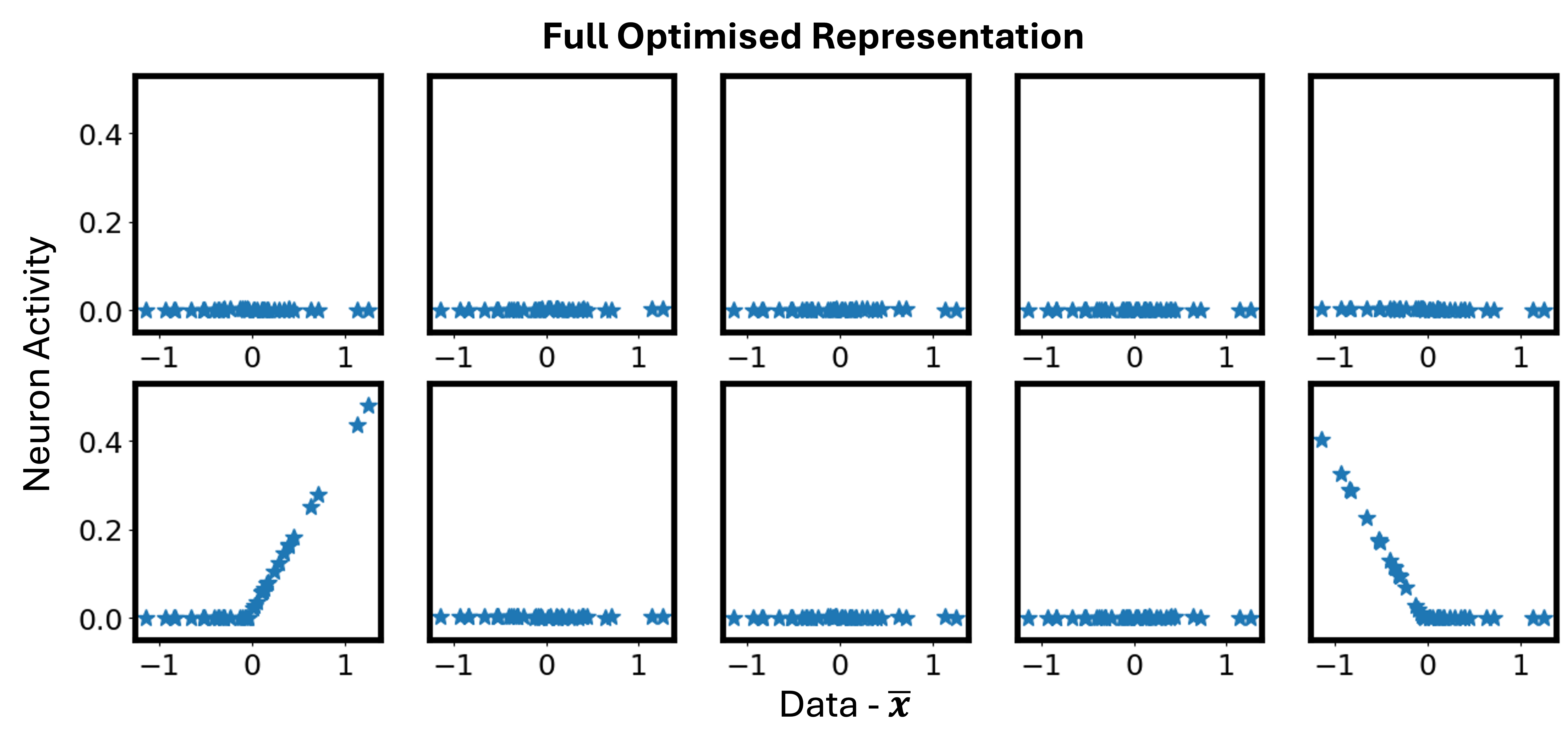}
    \caption{We optimise ten nonnegative neurons to encode a scalar variable by minimising \cref{eq:only_Z_dictionary_learning}. Only two neurons have non-zero activity and their functional form matches the prediction, \cref{fig:dense_antipodal}. Somehow optimisation selects this solution from amongst the degenerate set of optimal solutions.}
    \label{fig_app:full_dense_antipodal}
\end{figure}

We empirically test these claims by optimising a nonnegative encoding to minimise \cref{eq:only_Z_dictionary_learning}. We input a scalar and run gradient descent on $\mZ$ subject to nonnegativity. \Cref{fig:dense_antipodal} shows the comparison of the two non-zero neurons to the predictions from \cref{sec_app:precise_dense_antipodal_representation}. We include the full 10 neuron representation we optimised to illustrate how cleanly the representation uses only two neurons, \cref{fig_app:full_dense_antipodal}. The choice of using two neurons, rather than another of the degenerately optimal solutions, somehow results from optimisation.

\newpage

\section{Wide-Dictionary Setting}\label{app_sec:more_neurons_than_datapoints}

In this section we consider the structure of the optima when there is no constraint on the number of neurons. In this case the problem is convex,~\cref{app_sec:Q_reformulations}. \citet{bach2008convex} study a related problem and find that one-neuron-per-datapoint is a global minima . However, this problem has a slight non-triviality: we saw in \cref{sec_app:dense_encodings} that no matter how many neurons we added, the optimal solution could be achieved using at most 2 neurons. Here we show that indeed, one-neuron-per-datapoint is a global minima, but further, that the narrowest solution that achieves this global minima is governed by the number of rays from the optimal bias $\vb$ required to hit all the datapoints further than $\lambda$ from $\vb$.

Consider a representation that activates at most a single feature per datapoint:
\begin{equation}
    z_d^{[i]} = \begin{cases}
        a_d \geq 0\quad \text{if} \quad d = i \\
        0 \qquad \text{else}
    \end{cases}
\end{equation}
Let's consider the original version of the problem, \cref{eq_app:dictionary_learning_problem}. Then the optimal choice of $\vw_d$ is a unit vector aligned with $\bar\vx^{[i]}-\vb$, making the loss proportional to:
\begin{equation}
    \sum_i ||\bar\vx^{[i]}-\vb||^2\bigg(1-\frac{a^{[i]}}{||\bar\vx^{[i]}-\vb||}\bigg)^2 + 2\lambda\sum_i a^{[i]}
\end{equation}
Following the arguments in the main paper, \cref{sec:feature_splitting_limit}, there must be at most one active neuron per datapoint in the globally minimal solution whose dictionary vector is aligned along $\bar\vx^{[i]}-\vb$ for the $\bar\vx^{[i]}$ on which it is active. This constrains the active sets of each neuron to those along rays from $\vb$ and tells us that a global minima is exactly the set of ray solutions.

Optimising with respect to $a^{[i]}$ and taking into account the nonnegativity we get:
\begin{equation}
    a^{[i]} = \max(||\bar\vx^{[i]}-\vb|| - \lambda, 0)
\end{equation}
This tells us exactly that datapoints within $\lambda$ of $\vb$ are encoded with the zero vector, hence not contributing to the number of neurons needed to reach the minimal width global minima.

We find that $\vb$ can be found as the solution to the following optimisation problem:
\begin{equation}
    \vb = \arg \min_b \sum_i \ell_\lambda(||\bar\vx^{[i]} - \vb||) \quad \text{where} \quad \ell_\lambda(||\bar\vx^{[i]} - \vb||) = \begin{cases}
        ||\bar\vx^{[i]} - \vb||^2 \quad \qquad ||\bar\vx^{[i]} - \vb|| \leq \lambda \\
        2\lambda||\bar\vx^{[i]} - \vb|| - \lambda^2 \quad ||\bar\vx^{[i]} - \vb|| > \lambda
    \end{cases}
\end{equation}
Which is a Huber-like interpolation between a squared and linear penalty. In the limit of small $\lambda$ this becomes the geometric median \cref{eq:geometric_median}. 

\textbf{1D Dataset} We derived the representation of 1D data using a convoluted KKT argument, \cref{sec_app:dense_encodings}, however it can be much more simply seen here. We can shuffle activity amongst neurons pointing in the same direction without changing the loss. For 1D data there are at most two directions. Shuffling activity so that only two neurons are active, or one if possible, we recover the same solution.

\end{document}